\documentclass[12pt]{article}

\usepackage{scicite}
\usepackage{times}
\usepackage{amsmath}
\usepackage{amssymb}
\usepackage{bm}
\usepackage{dutchcal}
\usepackage{dsfont}
\usepackage{url}
\usepackage{makecell}
\usepackage{graphicx}
\usepackage[font=small]{caption}
\RequirePackage{jabbrv}

\DeclareMathOperator{\var}{var}
\newcommand{\lr}[1]{\left(#1\right)}
\newcommand{\lrs}[1]{\left[#1\right]}

\newenvironment{sciabstract}{%
\begin{quote} \bf}
{\end{quote}}

\newcounter{lastnote}

\title{} 

\author
{\LARGE{Tuna and billfish larval distributions in a warming ocean}\\
\\
Hirotaka Ijima,$^{1\ast}$ Marko Jusup,$^{1}$\\
\\
\normalsize{$^{1}$Fisheries Resources Institute, Japan Fisheries Research and Education Agency,}\\
\normalsize{Yokohama 236-8648, Japan}\\
\\
\normalsize{$^\ast$Corresponding author: ijima\_hirotaka69@fra.go.jp.}
}

\date{}

\begin{document} 

\baselineskip24pt

\maketitle 

\begin{sciabstract}
Tuna and billfish are charismatic pelagic fishes attracting considerable scientific attention due to their ecophysiological and socioeconomic importance. However, the knowledge of their basin-wide spawning and larval habitats, especially in a warming ocean, is limited. This knowledge gap undermines effective fishery management by introducing recruitment uncertainty, which makes population dynamics unpredictable. We fill the gap with a parsimonious geostatistical species-distribution model trained on the largest available dataset on tuna and billfish larvae in the Pacific Ocean. The model reveals (i) the basin-wide seasonal larval distributions over the reference period 1960-85, (ii) the expected impact of ongoing ocean warming on these distributions, and (iii) the biogeochemical factors, such as pH, phosphate concentration, and sea-surface height, that shape the larval habitat. Our findings make a quantum leap in understanding the ecophysiology of tuna and billfish, providing valuable information for future conservation efforts.
\end{sciabstract}

\section*{Introduction}

Tuna and billfish are known for their extraordinary physique that captivates public and scientific imagination alike. Tuna, for example, have motivated a large body of work on their swimming~\cite{watanabe2015comparative, shadwick2008thunniform, dewar1994studies} and metabolic~\cite{estess2014bioenergetics, jusup2014simple, jusup2011full, korsmeyer2001tuna, korsmeyer1996aerobic} performance, inspiring even the design of biomimetic underwater autonomous vehicles~\cite{wainwright2020tunas}. The ability to perform at high levels enables tuna and billfish to populate all oceans and migrate over extreme distances~\cite{watanabe2015comparative, reygondeau2012biogeography, graham2004tuna}, as well as occupy the ecological niche of apex marine predators~\cite{watanabe2015comparative, kitchell1999keystone, olson1986apex}.

The ecophysiology of tuna and billfish underpins their socioeconomic status. Their muscular bodies and cosmopolitan distribution make them a highly sought-after catch. Tuna fisheries collectively achieve a high monetary value~\cite{mckinney2020netting, guillotreau2017local}, whereas billfish are a prised gamefish~\cite{holland1998ecotourism}. The global exploitation of tuna and billfish has raised concerns about overfishing and the sustainability of wild stocks~\cite{collette2011high, juan2011global}. To address these concerns, major fishing powers are coordinating management and conservation efforts through the operations of regional fisheries management organisations (RFMOs), with the goal of establishing sustainable tuna and billfish fisheries.

An increasingly complete understanding of fish ecophysiology is essential for effective management and conservation of wild stocks~\cite{horodysky2016fisheries}. Ecophysiology provides insight into how functional traits contribute to fisheries-driven population risks~\cite{murua2017fast, juan2015population} and can help identify targets for monitoring that simplify population health assessments~\cite{haberle2023fish, marn2020quantifying, ijima2019effects}. Reflecting its importance, the ecophysiology of tuna and billfish has been continually studied since at least the 1960s~\cite{rossifanelli1960oxygen, graham1975heat, barkley1978skipjack, block2001tuna, kitagawa2015biology}. These efforts notwithstanding, important knowledge gaps persist, particularly with regards to basin-wide spawning and larval habitats in a warming ocean~\cite{llopiz2015global, reglero2014worldwide}. A consequence is that tuna and billfish stock assessments face substantial recruitment uncertainty, driving large population fluctuations~\cite{shelton2011fluctuations}, and thus undermining effective fishery management.

Related research has pursued a wide range of goals using diverse methodological approaches. Stable-isotope analysis has been employed to trace larval natal origins~\cite{hane2022natal, wells2021nursery}, histological evaluations to identify spawning grounds~\cite{schaefer2019spatiotemporal}, controlled experiments to determine optimal conditions for larval development~\cite{kim2015effect, wexler2011temperature}, DNA analysis to improve larval identification and pinpoint new spawning sites~\cite{hyde2006central}, and electronic tagging to record behaviours around known spawning areas~\cite{fujioka2021habitat}. However, these studies mostly focus on a single species or geographic region, which may lead to inconsistencies, as exemplified by conflicting findings on the temperature tolerance of yellowfin-tuna larvae~\cite{kim2015effect, wexler2011temperature}. This highlights the challenges of integrating fragmented data into a cohesive understanding.

\begin{figure}[!t]
  \centering
  \includegraphics[width=1\textwidth]{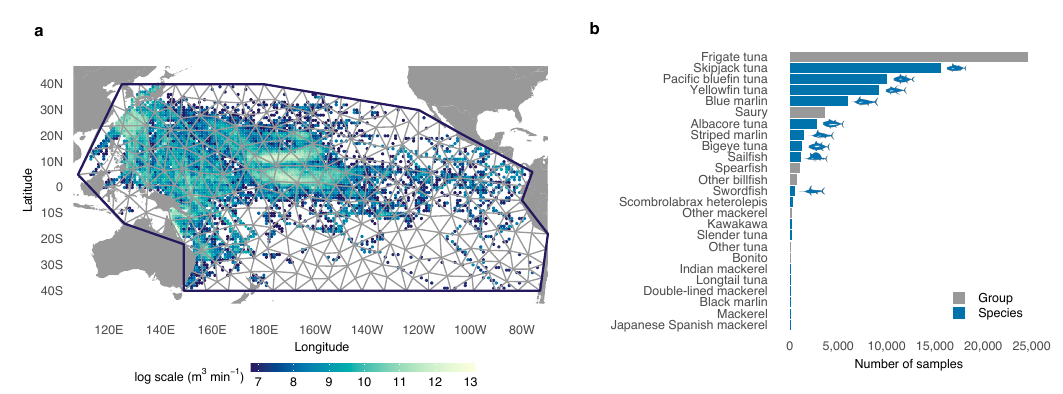}
  \caption{Larval-survey dataset. (a) The plot displays data points from the reference period 1960-85. Each data point includes records on geolocation, larval species, effort (in m$^3$\,min$^{-1}$), and sea surface temperature. The plot also shows a Delaunay triangulation of the area encompassing all larval-survey locations. The triangulation, comprising 277 nodes connected by edges with a maximum length of 1,250\,km, defines the domain over which our geostatistical species-distribution model operates. (b) Larval surveys observed a total of 24 individual species or multi-species groups. We focused our analyses on nine tuna and billfish species that were identified with high confidence. Ordered by the number of samples, these species are: skipjack tuna (\textit{Katsuwonus pelamis}), Pacific bluefin tuna (\textit{Thunnus orientalis}), yellowfin tuna (\textit{Thunnus albacares}), blue marlin (\textit{Makaira nigricans}), albacore tuna (\textit{Thunnus alalunga}), striped marlin (\textit{Kajikia audax}), bigeye tuna (\textit{Thunnus obesus}), sailfish (\textit{Istiophorus platypterus}), and swordfish (\textit{Xiphias gladius}).}
  \label{fig:dataset}
\end{figure}

We adopt a broader perspective using geostatistical species-distribution modelling to extract new fundamental and actionable information from a larval-survey dataset collected across the Pacific Ocean from 1960 to 1985 (Figure~\ref{fig:dataset}a). Specifically, the goals of the study were threefold: (i) to reconstruct basin-wide seasonal tuna and billfish larval distributions for the reference period 1960-85, (ii) to predict the expected changes to those distributions due to ongoing ocean warming, and (iii) to identify biogeochemical variables that are likely to shape larval abundance.

From a methodological standpoint, the larval-survey dataset, previously described in the literature~\cite{buenafe2022global, nishikawa1985average}, includes records of species, geolocation, effort (in m$^3$\,min$^{-1}$), and sea-surface temperature (SST). Our analysis focused on nine tuna and billfish species identified with confidence, out of the 24 species or multi-species groups documented in the dataset (Figure~\ref{fig:dataset}b). The modelling process involved hypothesising and fitting candidate models, followed by selecting the model that best balances explanatory power and mathematical parsimony (Supporting Table~\ref{tab:modelselection}; see also Supporting Figures~\ref{fig:fittedparams1}--\ref{fig:fittedparams5}). Careful attention was given to avoiding common statistical issues such as overfitting and model misspecification. These efforts were supported by diagnostic tests, which confirmed the robustness of the selected model (Supporting Figures~\ref{fig:diagnostics}--\ref{fig:diagnostics2}). Further details on methodology are provided in the Methods section.

\section*{Results}\label{sec:res}

\begin{figure}[!t]
  \centering
  \includegraphics[width=1\textwidth]{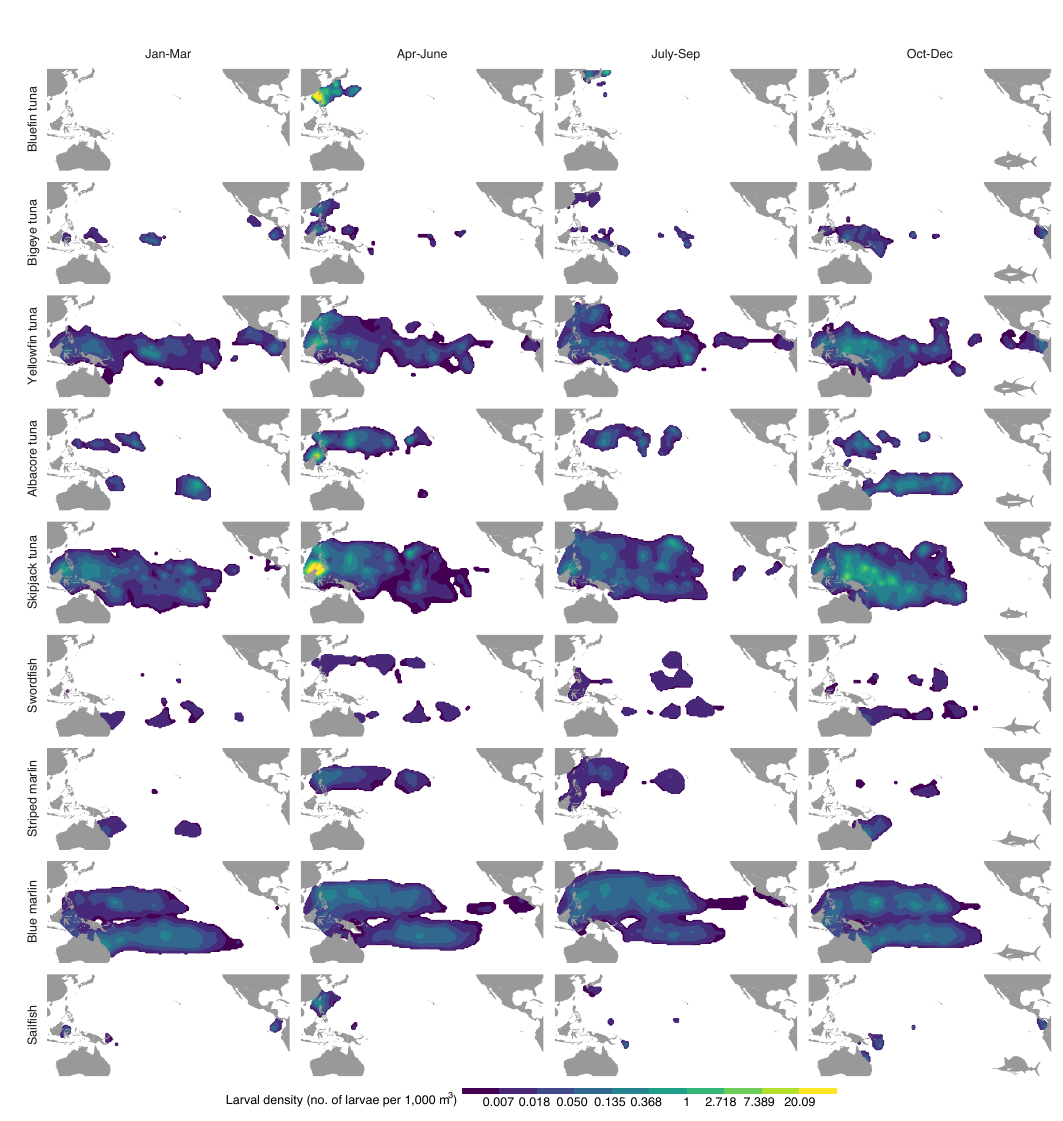}
  \caption{Seasonal tuna and billfish larval densities across the Pacific for the reference period 1960-85. In qualitative terms, we differentiate between (i) large-sized contiguous distributions of yellowfin-tuna, skipjack-tuna, and blue-marlin larvae, (ii) mid- to small-sized patchy distributions of bigeye-tuna, albacore-tuna, swordfish, striped-marlin, and sailfish larvae, and (iii) congregated distribution of Pacific bluefin-tuna larvae. Contiguous larval habitats are centred around the equator, with limited seasonal variability mirrored in the northward (southward) pull during northern (southern) warm months. Patchier larval habitats are subject to more pronounced seasonal changes, especially away from the equator where larvae tend to appear only during warm months. The Pacific bluefin-tuna larvae is exclusive to the northeastern Pacific in spring and, to a lesser degree, summer.}
  \label{fig:larvaldensities}
\end{figure}

The spatial and seasonal variations of larval distributions during the reference period 1960-85 exhibit three qualitative patterns (Figure~\ref{fig:larvaldensities}). The larvae of yellowfin tuna, skipjack tuna, and blue marlin are widely distributed around the equatorial Pacific. These distributions experience seasonal changes with a northward shift during boreal warm months and a southward shift during austral warm months. The larvae of bigeye tuna, albacore tuna, swordfish, striped marlin, and sailfish occupy medium to small patchy areas in the Pacific. These areas undergo substantial seasonal changes, particularly away from the equator where the presence or absence of larvae is mostly correlated with warm or cold months. The larvae of Pacific bluefin tuna are unique. They gather exclusively in the northwestern Pacific during boreal spring and summer.

\begin{figure}[!t]
  \centering
  \includegraphics[width=1\textwidth]{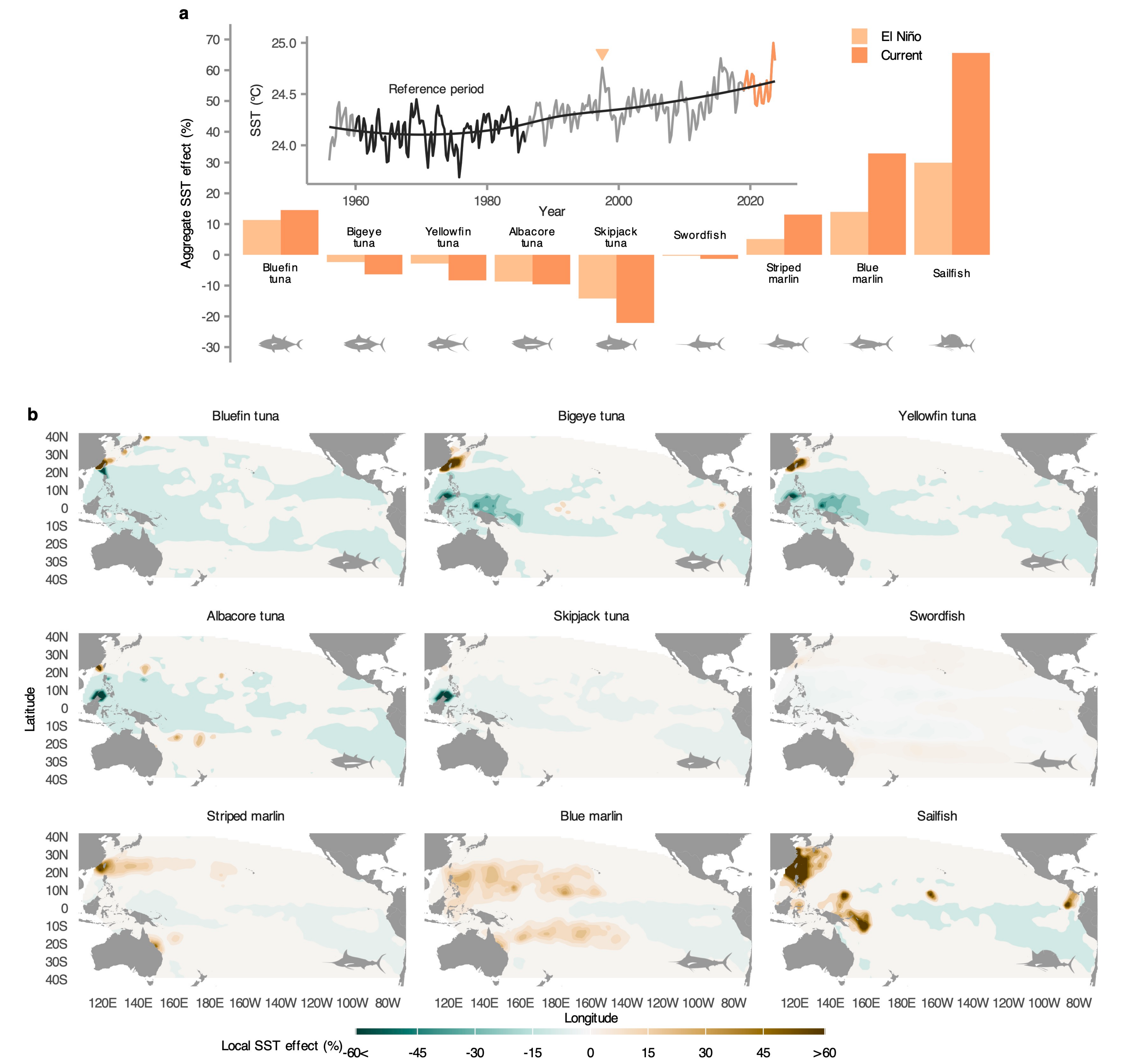}
  \caption{The effects of rising sea-surface temperature (SST) on tuna and billfish larval densities. (a) SST in the Pacific Ocean has risen in recent decades. We used the 1960-85 period as a reference for building our geostatistical species-distribution model. Relative to this reference period, the model predicts the basin-wide expected change in larval density during the major El Ni\~{n}o event of 1997-98 and under current conditions (2019-23). Bigeye tuna, yellowfin tuna, albacore tuna, and skipjack tuna are negatively impacted. The impact on swordfish is marginal. Pacific bluefin tuna, striped marlin, blue marlin, and sailfish are positively impacted. (b) The model also makes geospatially resolved predictions in terms of the local expected percentage change in larval density, displayed here for the current conditions. Brown and green shades respectively indicate areas where we can expect more or fewer larvae. The same predictions for the El Ni\~{n}o year are shown in Supporting Figure~\ref{fig:spatiallyresolved2}.}
  \label{fig:spatiallyresolved}
\end{figure}

The rising SST trend in the Pacific (Figure~\ref{fig:spatiallyresolved}a, inlet) is concerning given its potential to displace marine species~\cite{garcia2020global, oremus2020governance, burrows2019ocean, garcia2016climate, burrows2014geographical}. Our analysis indicates (Figure~\ref{fig:spatiallyresolved}a) that a negative aggregate effect should be expected for four tuna species (bigeye, yellowfin, albacore, and skipjack). Swordfish remain nearly unaffected, whereas a positive aggregate effect should be expected for Pacific bluefin tuna and three billfish species (striped marlin, blue marlin, and sailfish). The reason Pacific bluefin stands apart from other tuna species can be traced to a productivity-boosting increase in SST in the northern regions of its concentrated larval habitat, underscoring the importance of a geospatially resolved analysis (Figure~\ref{fig:spatiallyresolved}b). Ocean temperature trends vary by location, and what counts is the trend at locations where larvae are likely to be found.

\begin{figure}[!t]
  \centering
  \includegraphics[width=0.75\textwidth]{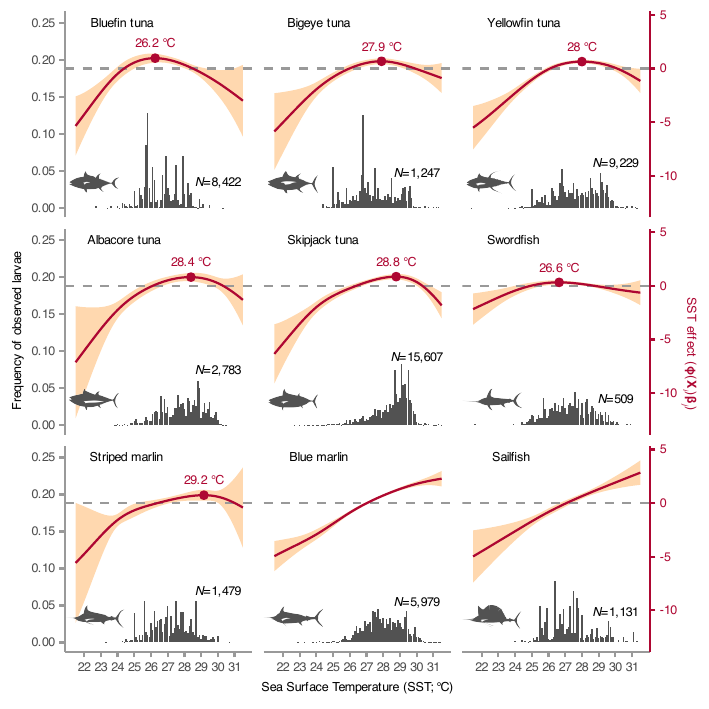}
  \caption{Temperature-response curves for tuna and billfish larvae in the Pacific Ocean. The curves show how larval densities change with sea surface temperature (SST) relative to a location's baseline. Curve envelopes represent the accompanying 95\,\% confidence intervals. Red-coloured curve maxima reveal SST at which larval densities peak, whereas grey-coloured horizontal dashed lines delineate positive from negative SST influence. Curve maxima could not be identified for blue marlin and sailfish. Histograms and the numbers accompanying them respectively show the temperature distributions of larval data and sample sizes.}
  \label{fig:ssteffects}
\end{figure}

The temperature-response curves for tuna and billfish larvae (Figure~\ref{fig:ssteffects}) provide further insight into the effects of rising SST. For seven of the nine species studied, the curves show a narrow SST range over which temperature boosts larval production beyond the baseline set by ecophysiology and other biogeochemical factors. Each of these seven species exhibits an optimal SST at which larval densities peak. Current SST is exceeding the optimum in key habitats of the four tuna species predicted to be negatively affected by temperature. In contrast, Pacific bluefin tuna and striped marlin benefit from SST approaching the optimum in their key habitats. Swordfish larvae turn out to be relatively insensitive to ongoing SST changes. The two remaining species, blue marlin and sailfish, show only increasing larval production with rising SST. This is likely a modelling artefact due to inconsistent presence of larvae in warm equatorial areas, to which the model then assigns a low baseline that is overcome by high temperature.

\begin{figure}[!t]
  \centering
  \includegraphics[width=1\textwidth]{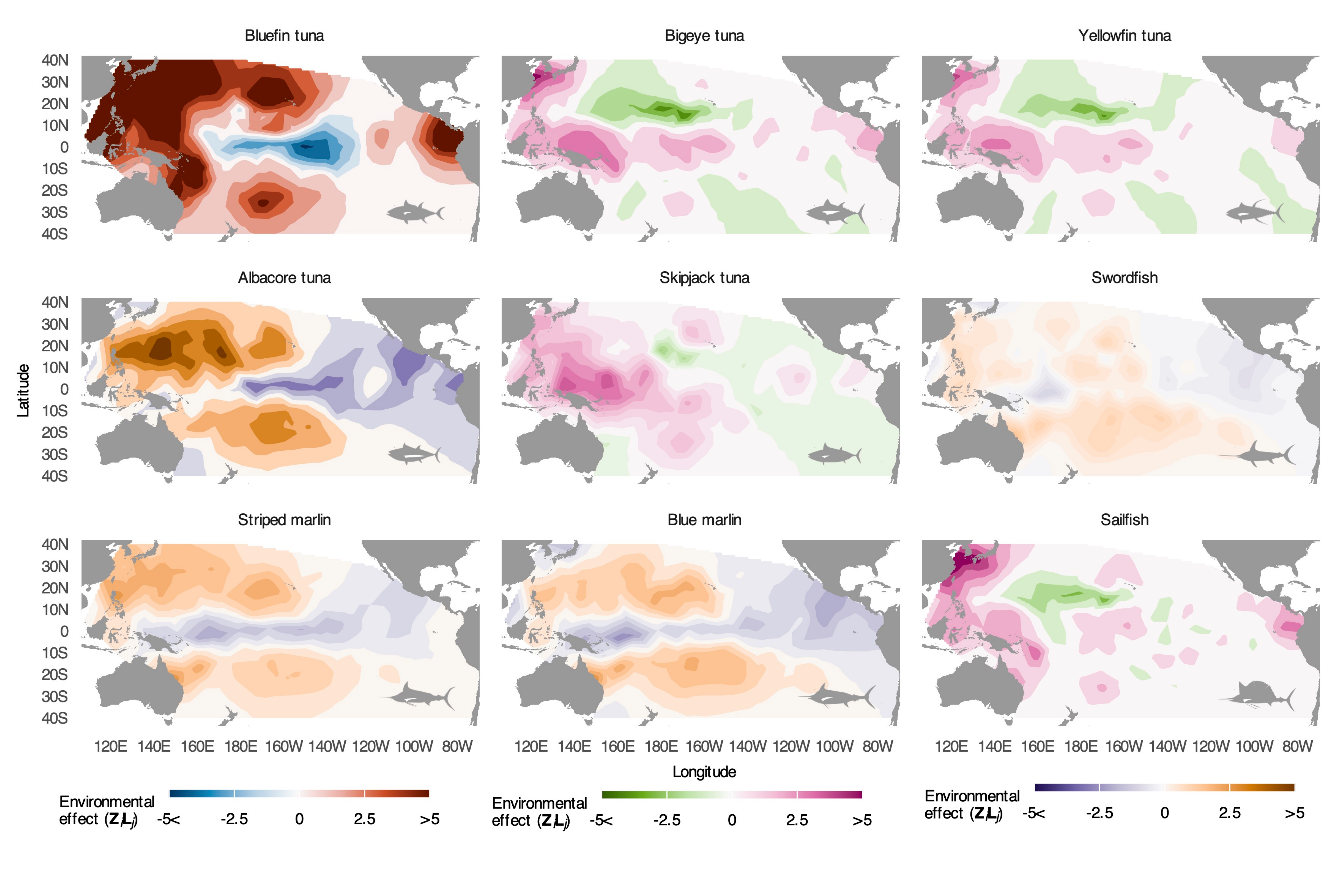}
  \caption{Potential of the oceanic environment excluding sea-surface temperature to support tuna and billfish larvae. The different colour schemes emphasise the visual similarity of environmental potentials, which upon a further quantitative analysis shown in Figure~\ref{fig:latentfactors}a, can be classified as either `tropical tuna-like' (bigeye tuna, yellowfin tuna, skipjack tuna, and sailfish) or `marlin-like' (striped marlin, blue marlin, swordfish, and albacore tuna), while Pacific bluefin tuna stands apart.}
  \label{fig:envpotential}
\end{figure}

\begin{figure}[!t]
  \centering
  \includegraphics[width=1\textwidth]{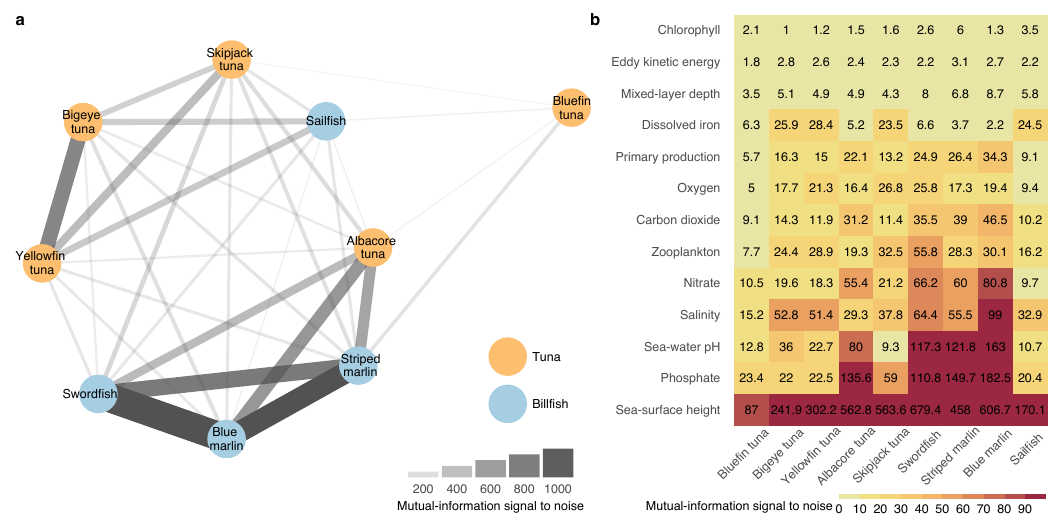}
  \caption{Characteristics and potential drivers of latent predictors. (a) A similarity network based on mutual-information signal to noise (visualised with both link width and shade) underpins the classification of environmental potentials shown in Figure~\ref{fig:envpotential}. Links with widths in the bottom 20th percentile are omitted. (c) Of the biogeochemical variables analysed, sea-surface-height variability, phosphate concentration, and pH show the strongest association with environmental potentials. We defined signal to noise as the mutual information between the environmental potential for a given species and a biogeochemical variable, divided by the same quantity when the variable is randomly reshuffled.}
  \label{fig:latentfactors}
\end{figure}

Our approach incorporates biogeochemical factors beyond SST into latent predictors, ensuring both parsimony and explanatory power while mitigating common statistical issues like misspecification and overfitting. The aggregate effect of all latent predictors reflects the potential of the environment net of SST to support tuna and billfish larvae (Figure~\ref{fig:envpotential}). Using mutual information to assess dataset similarities~\cite{reshef2011detecting}, we identified three types of environmental potentials, generally aligned with the phylogeny and life history of tuna and billfish species (Figure~\ref{fig:latentfactors}a). First, yellowfin, bigeye, skipjack, and surprisingly, sailfish share `tropical-tuna-like' environmental potentials. Second, striped marlin, blue marlin, swordfish, and surprisingly, albacore share `marlin-like' environmental potentials. Finally, Pacific bluefin tuna's environmental potential is distinct from others. The surprising environmental potentials for albacore tuna and sailfish suggest that similar adaptations may occur across phylogenetically distant species.

To pinpoint what shapes these environmental potentials, we analysed their associations with 13 geophysical, geochemical, and bioproductivity factors, including eddy kinetic energy, mixed-layer depth, sea-surface-height variability, dissolved oxygen, carbon dioxide, salinity, sea-water pH, dissolved iron, nitrate, phosphate, chlorophyll, primary production, and zooplankton (Supporting Figure~\ref{fig:biogeochemdata}). Using mutual information, we identified sea-water pH, phosphate, and sea-surface-height variability as the most relevant factors (Figure~\ref{fig:latentfactors}b), especially for marlin-like environmental potentials. Salinity also plays an important role, more so for tropical-tuna-like environmental potentials. The direction of these associations is consistent across all species. Accordingly, striped marlin, blue marlin, swordfish, and albacore tuna may be more susceptible to ocean acidification and more repelled by high-phosphate areas than yellowfin, bigeye, skipjack tuna, and sailfish. Elevated sea surface height appears universally beneficial for larval production.

\section*{Discussion}\label{sec:disc}

Our aim was to expand the ecophysiological understanding of tuna and billfish with emphasis on mapping larval habitats and quantifying the response of larvae to ongoing ocean warming. We achieved this by developing a geostatistical species-distribution model with high explanatory power, and training the model with the most comprehensive dataset on tuna and billfish larvae in the Pacific Ocean to date~\cite{buenafe2022global, nishikawa1985average}. Data availability notwithstanding, ours is among the first studies of its kind, arguably due to recent advancements in geostatistical species-distribution modelling~\cite{thorson2015spatial} and the computational resources required to generate basin-wide results. We used a workstation with 128 CPU cores and 1\,TB of memory during the modelling process.

\begin{figure}[!t]
  \centering
  \includegraphics[width=1\textwidth]{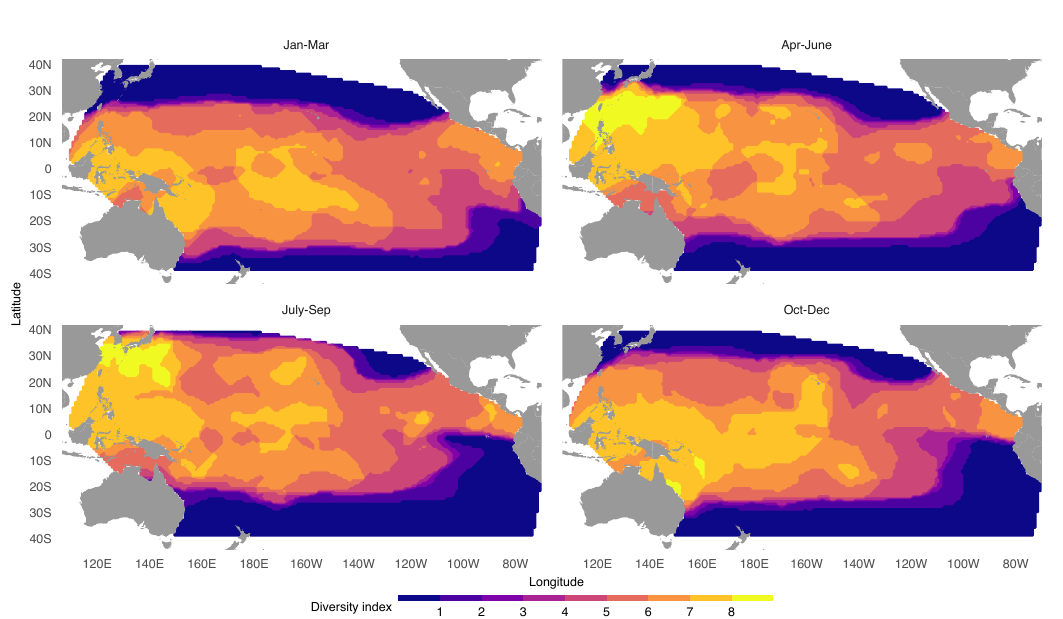}
  \caption{Larval diversity in the Pacific as inferred by our geostatistical species-distribution model. The western and central Pacific harbours noticeably more diversity than the eastern Pacific, with the overall distribution closely resembling the known distribution of sea-surface temperature~\cite{liu2019eastern}. Additionally, there is a noticeable diversity patch south of Japan during the boreal spring and summer, which coincides with the known distribution of sea-surface-height variability~\cite{kang2010eddy}.}
  \label{fig:larvaldiversity}
\end{figure}

Our approach may find many uses. Conservationists, for example, might use it to determine the distribution of larval diversity in the Pacific (Figure~\ref{fig:larvaldiversity}). This distribution reveals a distinctive divide in larval diversity across 140$^\circ$\,W longitude, and a hotspot of diversity south of Japan during boreal spring and summer. These patterns align with established distributions of SST~\cite{liu2019eastern} and sea-surface-height variability~\cite{kang2010eddy}, with the latter being the biogeochemical variable most closely linked to the model-generated environmental potentials and thereby larval densities.

The occurrence of larvae in our study is limited to the SST range 21.4-31.5\,$^\circ$C. Others have reported similar findings in the field~\cite{kitagawa2010restricted, hyde2006central} and in laboratory experiments~\cite{fujioka2024influence, margulies2007spawning}. It is well-known, however, that the temperature tolerance range for larvae is narrower than that for adults~\cite{pons2017effects}, which is why SST is considered a limiting factor for tuna and billfish reproduction. Sustained temperatures above 30\,$^\circ$C in critical areas could threaten reproductive activity.

Following the large-scale larval survey by Nishikawa et al.~\cite{nishikawa1985average}, subsequent studies have been spatially and temporally limited. Many confirm that spawning likely occurs within the spatial range predicted herein~\cite{hernandez2019evidence}. Some larval surveys and stable-isotope analyses, however, suggest that spawning may be occurring beyond our predicted range, particularly around Hawaii and along the eastern Pacific coast~\cite{wells2021nursery, armas1999confirmation}. We may have underestimated larval presence in these areas because effort of the original survey was low there. This underestimation, along with the failure to detect optimal SST values for blue marlin and sailfish, underscores the limitations of even the largest available larval dataset, and highlights the need for new basin-wide larval surveys.

Although there are no immediate solutions for the data and methodological limitations, ensuring the robustness of results under these constraints is essential. To this end, we appended previously mentioned model diagnostics (Supporting Figures~\ref{fig:diagnostics}--\ref{fig:diagnostics2}) with multiple sensitivity analyses, retaining data for the four most heavily sampled species while omitting one or more of the other five. The expected larval distributions for the reference period 1960-85 remain unaffected by these changes (Supporting Figure~\ref{fig:sensitivity}). We also examined the stability of temperature-response curves under various modelling and input scenarios, including replacing splines with parabolas, and found consistent results (Supporting Figure~\ref{fig:sensitivity2}). Notably, the model with parabolas enforced optimal SST estimates for blue marlin and sailfish, but the values exceeded 30\,$^\circ$C, challenging biological plausibility.

The SST effects may be amplified or suppressed by other biogeochemical variables. We found that sea-water pH correlates with lower larval densities as acidity increases. Increased acidity damages the internal organs of hatchlings, leading to higher larval mortality rates~\cite{wexler2023effect, frommel2016ocean}. Given that ocean warming is driven by rising atmospheric CO$_2$ concentrations, which also contribute to ocean acidification~\cite{jacobson2005studying}, substantial uncertainty surrounds even the future of the four species potentially benefiting from elevated SST. Notably, four of the nine species studied are at risk from both rising SST and declining sea-water pH, creating a double jeopardy for their well-being.

We also found that higher phosphate levels correlate with reduced larval densities. The larval diversity hotspot south of Japan coincides with one of the most phosphate-deficient regions in the Pacific Ocean~\cite{martiny2019biogeochemical}. This is particularly intriguing because phosphorus is widely regarded as the limiting nutrient for primary production~\cite{tyrrell1999relative}, despite recent research suggesting a more complex interaction~\cite{martiny2019biogeochemical}. Some studies propose that tuna and billfish larvae may exhibit reduced reliance on primary production due to larval cannibalism~\cite{reglero2011cannibalism}, with hatching in areas of low primary production potentially offering protection from predators~\cite{bakun2003environmental}. However, a clear causal relationship between phosphate availability and larval density remains to be established.

Interestingly, the areas of high larval diversity we identified closely resemble regions with high zooplankton diversity~\cite{rutherford1999environmental}. Before transitioning to a piscivorous diet, tuna and billfish larvae prey on zooplankton, exhibiting highly selective feeding behaviour~\cite{llopiz2015global}. Understanding how this selective zooplankton predation facilitates adaptation to low-productivity regions remains an unresolved ecological question.

Finally, we link higher sea-surface-height variability to increased larval densities. The impact of sea-surface height on tuna and billfish larval abundance has been previously documented~\cite{tawa2020fine, rooker2012distribution}, but the underlying causality remains as elusive as that for phosphorus. The arguments connecting negative sea-surface-height anomalies with upwelling and primary production~\cite{rooker2012distribution} are contradicted by our result showing no effect of chlorophyll concentration on larval densities.

Overall, causal relationships between biogeochemical variables and larval densities remain vague, presenting a number of crucial research questions for future studies. A natural starting point to pursuing these questions would be to consider biogeochemical variables emphasised by mutual information. One should keep in mind, however, that mutual information is akin to cross-correlation in that both fall short of guaranteeing the presence of underlying causality even between two strongly associated quantities.

\section*{Methods}\label{sec:meth}

\paragraph*{Model formulation.} We conducted the present study to add value to the largest available dataset on tuna and billfish larvae in the Pacific Ocean by extracting from the dataset new fundamental and actionable information on larval ecophysiology. To that end, we devised a parsimonious geostatistical species-distribution model with high explanatory power~\cite{warton2015so}. The model was built in three distinct steps.

First, we assumed that larval density at geolocation $\mathbf{r}_i$ and time $t_i$ for species $j$ is a random variable following the Tweedie distribution
\begin{equation}
y_{ij} \sim \mathcal{Tw}(\mu_{ij}, p_j, \phi_j), 1<p_j<2 \land \phi_j>0,
\end{equation}
where $\mu_{ij}$ is the mean, $p_j$ is the power parameter, and $\phi_j$ is the dispersion parameter. All three parameters jointly determine the distribution's variance $\var[y_{ij}] = \phi_j\mu_{ij}^{p_j}$. The Tweedie distribution is characterised by the non-negative real support $[0,+\infty)$ on which there is a probability mass concentrated at zero, followed by an exponentially decreasing profile away from zero. These characteristics naturally fit applications such as ours, in which many zero measurements are accompanied by occasional positive outcomes whose likelihood declines with size. 

The second step in model building was to represent the mean larval density as
\begin{equation}
\ln{\mu_{ij}} = \beta_j^0 + \bm{\Phi}(\mathbf{X}_i)\bm{\beta}_j + \mathbf{Z}_i\mathbf{L}_j,
\label{eq:meandensity}
\end{equation}
where $\beta_j^0$ is a species-specific constant term, $\bm{\Phi}=\bm{\Phi}(\mathbf{X}_i)$ is a feature map that transforms the vector $\mathbf{X}_i$ of observed predictor-variable values at geolocation $\mathbf{r}_i$ and time $t_i$ into a row vector of features, and $\bm{\beta}_j$ is a column vector of model coefficients for species $j$. We considered two common types of features involving exponentiated or Gaussian-transformed observations (details below). The row vector $\mathbf{Z}_i$ is a latent (spatial) analogue of the vector $\mathbf{X}_i$, incorporating further mathematical structure relevant for modelling larval densities. The column vector $\mathbf{L}_j$ is a latent analogue of the vector $\bm{\beta}_j$ with components \smash{$l_j^f$} such that \smash{$l_j^f\ne 0$} if $f<j$, \smash{$l_j^f=1$} if $f=j$, and \smash{$l_j^f=0$} if $f>j$, where $f$ indexes latent predictors.

The final step in model building was to assign mathematical structure to the vector $\mathbf{Z}_i$. The components $z_i^f$ of this vector are interpolated using \smash{$z_i^f=\mathbf{A}(\mathbf{r}_i)\mathbf{U}^f(t_i)$}, where $\mathbf{A}(\mathbf{r}_i)$ is a row vector containing the barycentric coordinates of geolocation $\mathbf{r}_i$ relative to a chosen Delaunay triangulation of the study area~\cite{cameletti2013spatio}, and $t_i$ is the corresponding timestamp. The Delaunay triangulation covers the study area with a set of nodes, indexed by $k$, at which we keep track of the components $u_{k,t}^f$ of \smash{$\mathbf{U}^f(t)$}.

We assumed that the latent predictors are temporally and spatially auto-correlated. The former is captured by
\begin{equation}
u_{k,t}^f = \sum_{s=1}^{4} \rho_s^f u_{k,t-s}^f + \omega_{k,t}^f,
\end{equation}
where $\rho_s^f$ are auto-regressive coefficients that determine the temporal auto-correlation structure of the overall model. The remaining term $\omega_{k,t}^f$ represents a realisation at node $k$ and time $t$ of the Gaussian Markov random field $\bm{\omega}^f\sim\mathrm{GMRF}(0,\bm{\Sigma}^f)$ whose covariance matrix $\bm{\Sigma}^f$ stores the spatial auto-correlation structure of the model. To satisfy the Markovian property, the covariance between any two points distance $d$ apart is given by the Mat\'{e}rn-type covariance function
\begin{equation}
M_\mathrm{cov}^f(d)=\frac{1}{4\pi}\frac{d}{\kappa^f\lr{\tau^f}^2}K_1\lr{\kappa^f d},
\end{equation}
where $K_1(\cdot)$ is the first-order modified Bessel function of the second kind, and $\kappa^f$ and $\tau^f$ are positive parameters.

\paragraph*{Implementation.} We estimated the model's parameters using the maximum-likelihood method. This method entails choosing the parameter values that maximise the log-likelihood of observing the dataset at hand. The log-likelihood is
\begin{align}
\log \mathcal{L} &= \sum\limits_i \sum\limits_j \log\, p\lr{y_{ij}\vert \beta_j^0, \bm{\beta}_j, \mathbf{L}_j, \ldots; \bm{\Sigma}^f} \nonumber\\
&= \sum\limits_i \sum\limits_j \log \int p\lr{y_{ij}\vert \ldots; \bm{\omega}^f} p\lr{\bm{\omega}^f\vert \bm{\Sigma}^f} \prod\limits_f\mathrm{d}\bm{\omega}^f,
\end{align}
where $p(\cdot\vert\cdot)$ signify probability density functions as per model definitions. Specifically, $p\lr{y_{ij}\vert \dots; \bm{\omega}^f}$ is the probability density function for the Tweedie distribution $\mathcal{Tw}(\mu_{ij}, p_j, \phi_j)$, whereas $p\lr{\bm{\omega}^f\vert \bm{\Sigma}^f}$ is the joint probability density function for a set of independent multivariate normal distributions with the covariance matrices $\Sigma^f$. The log-likelihood was, due to its complexity, maximised numerically with the R package TMB~\cite{kristensen2016tmb}.

Key modelling decisions involved selecting appropriate predictor datasets and feature maps. Sea-surface temperature (SST) was the only biogeochemical variable consistently measured during the original larval survey. Although we considered including other processed datasets from well-known oceanographic sources, we ultimately opted for a different analytic approach. We let the model isolate SST effects through the product $\bm{\Phi}(\mathbf{X}_i)\bm{\beta}_j$, while subsuming the influence of latent predictors into the product $\mathbf{Z}_i\mathbf{L}_j$, which quantifies the oceanic environment's potential to host tuna and billfish larvae depending on biogeochemical factors excluding SST. To identify the most influential biogeochemical factors, we assessed the shared information content between environmental potential and biogeochemical datasets using mutual information, a robust measure of both linear and non-linear associations~\cite{reshef2011detecting}. This approach sidestepped hidden uncertainties in processed datasets and complex dataset co-dependencies (e.g., multicollinearity), ultimately reducing the risks of misspecification and overfitting.

For the feature map $\bm{\Phi}(\cdot)$, we considered two common options. If we denoted a generic observed value with $x$, the first option was to exponentiate this value as if performing a Taylor expansion, $x\mapsto(x,x^2,\ldots)$. The second option was to follow the statistical function estimation using cubic splines~\cite{wood2020inference}, or Gaussians as their close approximation~\cite{wang1998scale}, in which case $x\mapsto(B_{1,3}(x),B_{2,3}(x),\ldots)$, where the indices respectively identify spline knots and order (here, 3 for cubic splines).

We interpreted model outputs as follows. Upon estimating the model's parameter vectors, Eq.~\eqref{eq:meandensity} could be treated as continuous in location $\mathbf{r}$ and time $t$, allowing us to rewrite $\mu_{ij}$ as $\mu_j(\mathbf{r},t)$. Time averaging over the reference period 1960-85 yielded the time-independent mean larval density, $\mu_{j,\mathrm{ref}}(\mathbf{r})$, representative of that period. Additionally, integrating over the study area, \smash{$M_{j,\mathrm{ref}}=\int{\mu_\mathrm{ref}(\mathbf{r})}\mathrm{d}^3\mathbf{r}$}, gave us the reference larval mass. The ratio \smash{$\frac{\mu_{j,\mathrm{ref}}(\mathbf{r})}{M_{j,\mathrm{ref}}}$} could then be interpreted as the probability density function of tuna and billfish larvae, leading to the expected percentage change in larval densities due to ocean warming at time $t$ as
\begin{equation}
\mathop{{}\mathbb{E}} \lrs{\Delta\%\mu_j(t)} = 100\,\%\times \int{\lr{\frac{\mu_j(\mathbf{r},t)}{\mu_{j,\mathrm{ref}}(\mathbf{r})}-1}\frac{\mu_{j,\mathrm{ref}}(\mathbf{r})}{M_{j,\mathrm{ref}}}}\mathrm{d}^3\mathbf{r}.
\end{equation}

To quantify diversity at location $\mathbf{r}$, we defined a diversity index
\begin{equation}
\mathrm{DI}(\mathbf{r}) = \sum\limits_j \mathds{1}_{\lbrace\mu_{j,\mathrm{ref}}(\mathbf{r}')>0\rbrace}(\mathbf{r}),
\end{equation}
where $\mathds{1}_A(\mathbf{r})$ is an indicator function such that $\mathds{1}_A(\mathbf{r})=1$ if $\mathbf{r}\in A$ and $\mathds{1}_A(\mathbf{r})=0$ otherwise. The quantity $\mathrm{DI}(\mathbf{r})$ is a simple counter of positive reference larval densities at a chosen location.

\paragraph*{Data availability.} The data on larval abundance are publicly accessible from Zenodo at \url{https://doi.org/10.5281/zenodo.6592148}. The data on sea-surface temperature (SST) used to predict the effects of ocean warming on tuna and billfish larvae are from the Hadley Centre Sea Ice and Sea Surface Temperature data set (HadISST) publicly accessible at \url{https://www.metoffice.gov.uk/hadobs/hadisst/}. The SST data measured during the original larval survey and used for model learning are a property of the government of Japan; requests for access should be directed at the corresponding author. As an alternative, we confirmed that the HadISST data for the reference period correlates strongly with the measured SST data, and is sufficient to replicate the study. The results of the replicability run are available at \url{https://doi.org/10.17605/OSF.IO/42HM8}. The data on mixed layer depth and salinity are publicly accessible from The World Ocean Atlas (WOA18) at \url{https://www.ncei.noaa.gov/access/world-ocean-atlas-2018/}. The data on sea surface height relative to geoid are publicly accessible from the NCEP Global Ocean Data Assimilation System (GODAS) at \url{https://psl.noaa.gov/data/gridded/data.godas.html}. The data on surface partial pressure of carbon dioxide in sea water, mole concentration of dissolved molecular oxygen in sea water, mass concentration of chlorophyll a in sea water, mole concentration of phosphate in sea water, sea water pH reported on total scale, and mole concentration of nitrate in sea water are publicly accessible from the Copernicus Marine Service Information at \url{https://marine.copernicus.eu/access-data}. The data on eddy kinetic energy are a part of the Archiving, Validation and Interpretation of Satellite Oceanographic (AVISO) data publicly accessible at \url{https://www.aviso.altimetry.fr/en/data/data-access.html}.

\paragraph*{Code availability.} The code developed for this analysis is publicly accessible from the Open Science Framework (OSF) at \url{https://doi.org/10.17605/OSF.IO/42HM8}.


\begin{thebibliography}{10}
\urlstyle{rm}
\expandafter\ifx\csname url\endcsname\relax
  \def\url#1{\texttt{#1}}\fi
\expandafter\ifx\csname urlprefix\endcsname\relax\def\urlprefix{URL }\fi
\expandafter\ifx\csname doiprefix\endcsname\relax\def\doiprefix{DOI: }\fi
\providecommand{\bibinfo}[2]{#2}
\providecommand{\eprint}[2][]{\url{#2}}

\bibitem{watanabe2015comparative}
\bibinfo{author}{Watanabe, Y.~Y.}, \bibinfo{author}{Goldman, K.~J.},
  \bibinfo{author}{Caselle, J.~E.}, \bibinfo{author}{Chapman, D.~D.} \&
  \bibinfo{author}{Papastamatiou, Y.~P.}
\newblock \bibinfo{journal}{\bibinfo{title}{Comparative analyses of
  animal-tracking data reveal ecological significance of endothermy in
  fishes}}.
\newblock {\emph{\JournalTitle{Proc. Natl. Acad. Sci. USA}}}
  \textbf{\bibinfo{volume}{112}}, \bibinfo{pages}{6104--6109}
  (\bibinfo{year}{2015}).

\bibitem{shadwick2008thunniform}
\bibinfo{author}{Shadwick, R.~E.} \& \bibinfo{author}{Syme, D.~A.}
\newblock \bibinfo{journal}{\bibinfo{title}{Thunniform swimming: muscle
  dynamics and mechanical power production of aerobic fibres in yellowfin tuna
  (\textit{Thunnus albacares})}}.
\newblock {\emph{\JournalTitle{J. Exp. Biol.}}} \textbf{\bibinfo{volume}{211}},
  \bibinfo{pages}{1603--1611} (\bibinfo{year}{2008}).

\bibitem{dewar1994studies}
\bibinfo{author}{Dewar, H.} \& \bibinfo{author}{Graham, J.}
\newblock \bibinfo{journal}{\bibinfo{title}{Studies of tropical tuna swimming
  performance in a large water tunnel---energetics}}.
\newblock {\emph{\JournalTitle{J. Exp. Biol.}}} \textbf{\bibinfo{volume}{192}},
  \bibinfo{pages}{13--31} (\bibinfo{year}{1994}).

\bibitem{estess2014bioenergetics}
\bibinfo{author}{Estess, E.~E.}, \bibinfo{author}{Coffey, D.~M.},
  \bibinfo{author}{Shimose, T.}, \bibinfo{author}{Seitz, A.~C.},
  \bibinfo{author}{Rodriguez, L.} \emph{et~al.}
\newblock \bibinfo{journal}{\bibinfo{title}{Bioenergetics of captive pacific
  bluefin tuna (\textit{Thunnus orientalis})}}.
\newblock {\emph{\JournalTitle{Aquaculture}}} \textbf{\bibinfo{volume}{434}},
  \bibinfo{pages}{137--144} (\bibinfo{year}{2014}).

\bibitem{jusup2014simple}
\bibinfo{author}{Jusup, M.}, \bibinfo{author}{Klanj{\v{s}}{\v{c}}ek, T.} \&
  \bibinfo{author}{Matsuda, H.}
\newblock \bibinfo{journal}{\bibinfo{title}{Simple measurements reveal the
  feeding history, the onset of reproduction, and energy conversion
  efficiencies in captive bluefin tuna}}.
\newblock {\emph{\JournalTitle{J. Sea Res.}}} \textbf{\bibinfo{volume}{94}},
  \bibinfo{pages}{144--155} (\bibinfo{year}{2014}).

\bibitem{jusup2011full}
\bibinfo{author}{Jusup, M.}, \bibinfo{author}{Klanj{\v{s}}{\v{c}}ek, T.},
  \bibinfo{author}{Matsuda, H.} \& \bibinfo{author}{Kooijman, S. A. L.~M.}
\newblock \bibinfo{journal}{\bibinfo{title}{A full lifecycle bioenergetic model
  for bluefin tuna}}.
\newblock {\emph{\JournalTitle{PLOS One}}} \textbf{\bibinfo{volume}{6}},
  \bibinfo{pages}{e21903} (\bibinfo{year}{2011}).

\bibitem{korsmeyer2001tuna}
\bibinfo{author}{Korsmeyer, K.~E.} \& \bibinfo{author}{Dewar, H.}
\newblock \bibinfo{journal}{\bibinfo{title}{Tuna metabolism and energetics}}.
\newblock {\emph{\JournalTitle{Fish Physiol.}}} \textbf{\bibinfo{volume}{19}},
  \bibinfo{pages}{35--78} (\bibinfo{year}{2001}).

\bibitem{korsmeyer1996aerobic}
\bibinfo{author}{Korsmeyer, K.~E.}, \bibinfo{author}{Dewar, H.},
  \bibinfo{author}{Lai, N.~C.} \& \bibinfo{author}{Graham, J.~B.}
\newblock \bibinfo{journal}{\bibinfo{title}{The aerobic capacity of tunas:
  adaptation for multiple metabolic demands}}.
\newblock {\emph{\JournalTitle{Comp. Biochem. Physiol.}}}
  \textbf{\bibinfo{volume}{113}}, \bibinfo{pages}{17--24}
  (\bibinfo{year}{1996}).

\bibitem{wainwright2020tunas}
\bibinfo{author}{Wainwright, D.~K.} \& \bibinfo{author}{Lauder, G.~V.}
\newblock \bibinfo{journal}{\bibinfo{title}{Tunas as a high-performance fish
  platform for inspiring the next generation of autonomous underwater
  vehicles}}.
\newblock {\emph{\JournalTitle{Bioinspir. Biomim.}}}
  \textbf{\bibinfo{volume}{15}}, \bibinfo{pages}{035007}
  (\bibinfo{year}{2020}).

\bibitem{reygondeau2012biogeography}
\bibinfo{author}{Reygondeau, G.}, \bibinfo{author}{Maury, O.},
  \bibinfo{author}{Beaugrand, G.}, \bibinfo{author}{Fromentin, J.~M.},
  \bibinfo{author}{Fonteneau, A.} \& \bibinfo{author}{Cury, P.}
\newblock \bibinfo{journal}{\bibinfo{title}{Biogeography of tuna and billfish
  communities}}.
\newblock {\emph{\JournalTitle{J. Biogeogr.}}} \textbf{\bibinfo{volume}{39}},
  \bibinfo{pages}{114--129} (\bibinfo{year}{2012}).

\bibitem{graham2004tuna}
\bibinfo{author}{Graham, J.~B.} \& \bibinfo{author}{Dickson, K.~A.}
\newblock \bibinfo{journal}{\bibinfo{title}{Tuna comparative physiology}}.
\newblock {\emph{\JournalTitle{J. Exp. Biol.}}} \textbf{\bibinfo{volume}{207}},
  \bibinfo{pages}{4015--4024} (\bibinfo{year}{2004}).

\bibitem{kitchell1999keystone}
\bibinfo{author}{Kitchell, J.~F.}, \bibinfo{author}{Boggs, C.~H.},
  \bibinfo{author}{He, X.} \& \bibinfo{author}{Walters, C.~J.}
\newblock \bibinfo{title}{Keystone predators in the central {P}acific}.
\newblock In \bibinfo{editor}{Mecklenburg, C.~W.} (ed.)
  \emph{\bibinfo{booktitle}{Ecosystem Approaches for Fisheries Management}},
  vol.~\bibinfo{volume}{16} of \emph{\bibinfo{series}{Lowell Wakefield
  Fisheries Symposium}}, \bibinfo{pages}{665--683}
  (\bibinfo{publisher}{University of Alaska Sea Grant College Program},
  \bibinfo{year}{1999}).

\bibitem{olson1986apex}
\bibinfo{author}{Olson, R.~J.} \& \bibinfo{author}{Boggs, C.~H.}
\newblock \bibinfo{journal}{\bibinfo{title}{Apex predation by yellowfin tuna
  (\textit{Thunnus albacares}): independent estimates from gastric evacuation
  and stomach contents, bioenergetics, and cesium concentrations}}.
\newblock {\emph{\JournalTitle{Can. J. Fish. Aquat. Sci.}}}
  \textbf{\bibinfo{volume}{43}}, \bibinfo{pages}{1760--1775}
  (\bibinfo{year}{1986}).

\bibitem{mckinney2020netting}
\bibinfo{author}{McKinney, R.}, \bibinfo{author}{Gibbon, J.},
  \bibinfo{author}{Wozniak, E.} \& \bibinfo{author}{Galland, G.}
\newblock \bibinfo{title}{Netting billions 2020: {A} global tuna valuation}.
\newblock \bibinfo{howpublished}{The Pew Charitable Trusts}
  (\bibinfo{year}{2020}).
\newblock \bibinfo{note}{Available at:
  \url{https://www.pewtrusts.org/-/media/assets/2020/10/nettingbillions2020.pdf}.}

\bibitem{guillotreau2017local}
\bibinfo{author}{Guillotreau, P.}, \bibinfo{author}{Squires, D.},
  \bibinfo{author}{Sun, J.} \& \bibinfo{author}{Compe{\'a}n, G.~A.}
\newblock \bibinfo{journal}{\bibinfo{title}{Local, regional and global markets:
  what drives the tuna fisheries?}}
\newblock {\emph{\JournalTitle{Rev. Fish Biol. Fish.}}}
  \textbf{\bibinfo{volume}{27}}, \bibinfo{pages}{909--929}
  (\bibinfo{year}{2017}).

\bibitem{holland1998ecotourism}
\bibinfo{author}{Holland, S.~M.}, \bibinfo{author}{Ditton, R.~B.} \&
  \bibinfo{author}{Graefe, A.~R.}
\newblock \bibinfo{journal}{\bibinfo{title}{An ecotourism perspective on
  billfish fisheries}}.
\newblock {\emph{\JournalTitle{J. Sustain. Tour.}}}
  \textbf{\bibinfo{volume}{6}}, \bibinfo{pages}{97--116}
  (\bibinfo{year}{1998}).

\bibitem{collette2011high}
\bibinfo{author}{Collette, B.~B.}, \bibinfo{author}{Carpenter, K.~E.},
  \bibinfo{author}{Polidoro, B.~A.}, \bibinfo{author}{Juan-Jord\'{a}, M.~J.},
  \bibinfo{author}{Boustany, A.} \emph{et~al.}
\newblock \bibinfo{journal}{\bibinfo{title}{High value and long life—double
  jeopardy for tunas and billfishes}}.
\newblock {\emph{\JournalTitle{Science}}} \textbf{\bibinfo{volume}{333}},
  \bibinfo{pages}{291--292} (\bibinfo{year}{2011}).

\bibitem{juan2011global}
\bibinfo{author}{Juan-Jord{\'a}, M.~J.}, \bibinfo{author}{Mosqueira, I.},
  \bibinfo{author}{Cooper, A.~B.}, \bibinfo{author}{Freire, J.} \&
  \bibinfo{author}{Dulvy, N.~K.}
\newblock \bibinfo{journal}{\bibinfo{title}{Global population trajectories of
  tunas and their relatives}}.
\newblock {\emph{\JournalTitle{Proc. Natl. Acad. Sci. USA}}}
  \textbf{\bibinfo{volume}{108}}, \bibinfo{pages}{20650--20655}
  (\bibinfo{year}{2011}).

\bibitem{horodysky2016fisheries}
\bibinfo{author}{Horodysky, A.~Z.}, \bibinfo{author}{Cooke, S.~J.},
  \bibinfo{author}{Graves, J.~E.} \& \bibinfo{author}{Brill, R.~W.}
\newblock \bibinfo{journal}{\bibinfo{title}{Fisheries conservation on the high
  seas: linking conservation physiology and fisheries ecology for the
  management of large pelagic fishes}}.
\newblock {\emph{\JournalTitle{Conserv. Physiol.}}}
  \textbf{\bibinfo{volume}{4}}, \bibinfo{pages}{cov059} (\bibinfo{year}{2016}).

\bibitem{murua2017fast}
\bibinfo{author}{Murua, H.}, \bibinfo{author}{Rodriguez-Marin, E.},
  \bibinfo{author}{Neilson, J.~D.}, \bibinfo{author}{Farley, J.~H.} \&
  \bibinfo{author}{Juan-Jord{\'a}, M.~J.}
\newblock \bibinfo{journal}{\bibinfo{title}{Fast versus slow growing tuna
  species: age, growth, and implications for population dynamics and fisheries
  management}}.
\newblock {\emph{\JournalTitle{Rev. Fish Biol. Fish}}}
  \textbf{\bibinfo{volume}{27}}, \bibinfo{pages}{733--773}
  (\bibinfo{year}{2017}).

\bibitem{juan2015population}
\bibinfo{author}{Juan-Jord{\'a}, M.~J.}, \bibinfo{author}{Mosqueira, I.},
  \bibinfo{author}{Freire, J.} \& \bibinfo{author}{Dulvy, N.~K.}
\newblock \bibinfo{journal}{\bibinfo{title}{Population declines of tuna and
  relatives depend on their speed of life}}.
\newblock {\emph{\JournalTitle{Proc. Royal Soc. B Biol. Sci.}}}
  \textbf{\bibinfo{volume}{282}}, \bibinfo{pages}{20150322}
  (\bibinfo{year}{2015}).

\bibitem{haberle2023fish}
\bibinfo{author}{Haberle, I.}, \bibinfo{author}{Bavcevic, L.} \&
  \bibinfo{author}{Klanj{\v{s}}{\v{c}}ek, T.}
\newblock \bibinfo{journal}{\bibinfo{title}{Fish condition as an indicator of
  stock status: {I}nsights from condition index in a food-limiting
  environment}}.
\newblock {\emph{\JournalTitle{Fish Fish.}}} \textbf{\bibinfo{volume}{n/a}},
  \doiprefix\url{10.1111/faf.12744} (\bibinfo{year}{2023}).

\bibitem{marn2020quantifying}
\bibinfo{author}{Marn, N.}, \bibinfo{author}{Jusup, M.},
  \bibinfo{author}{Kooijman, S.~A.} \& \bibinfo{author}{Klanj{\v{s}}{\v{c}}ek, T.}
\newblock \bibinfo{journal}{\bibinfo{title}{Quantifying impacts of plastic
  debris on marine wildlife identifies ecological breakpoints}}.
\newblock {\emph{\JournalTitle{Ecol. Lett.}}} \textbf{\bibinfo{volume}{23}},
  \bibinfo{pages}{1479--1487} (\bibinfo{year}{2020}).

\bibitem{ijima2019effects}
\bibinfo{author}{Ijima, H.}, \bibinfo{author}{Jusup, M.},
  \bibinfo{author}{Takada, T.}, \bibinfo{author}{Akita, T.},
  \bibinfo{author}{Matsuda, H.} \& \bibinfo{author}{Klanj{\v{s}}{\v{c}}ek, T.}
\newblock \bibinfo{journal}{\bibinfo{title}{Effects of environmental change and
  early-life stochasticity on pacific bluefin tuna population growth}}.
\newblock {\emph{\JournalTitle{Mar. Environ. Res.}}}
  \textbf{\bibinfo{volume}{149}}, \bibinfo{pages}{18--26}
  (\bibinfo{year}{2019}).

\bibitem{rossifanelli1960oxygen}
\bibinfo{author}{Rossi-Fanelli, A.} \& \bibinfo{author}{Antonini, E.}
\newblock \bibinfo{journal}{\bibinfo{title}{Oxygen equilibrium of haemoglobin
  from \textit{Thunnus thynnus}}}.
\newblock {\emph{\JournalTitle{Nature}}}
  \textbf{\bibinfo{volume}{186}}, \bibinfo{pages}{895--896}
  (\bibinfo{year}{1960}).

\bibitem{graham1975heat}
\bibinfo{author}{Graham, J.~B.}
\newblock \bibinfo{journal}{\bibinfo{title}{Heat exchange in the yellowfin tuna,
  \textit{Thunnus albacares}, and skipjack tuna, \textit{Katsuwonus pelamis}, and
  the adaptive significance of elevated body temperatures in scombrid fishes}}.
\newblock {\emph{\JournalTitle{Fish. Bull.}}}
  \textbf{\bibinfo{volume}{73}}, \bibinfo{pages}{219--229}
  (\bibinfo{year}{1975}).

\bibitem{barkley1978skipjack}
\bibinfo{author}{Barkley, R.~A.}, \bibinfo{author}{Neill, W.~H.}
  \& \bibinfo{author}{Gooding, R.~M.}
\newblock \bibinfo{journal}{\bibinfo{title}{Skipjack tuna, \textit{Katsuwonus pelamis},
  habitat based on temperature and oxygen requirements}}.
\newblock {\emph{\JournalTitle{Fish. Bull.}}}
  \textbf{\bibinfo{volume}{76}}, \bibinfo{pages}{653--662}
  (\bibinfo{year}{1978}).

\bibitem{block2001tuna}
\bibinfo{editor}{Block, B.~A.} \& \bibinfo{author}{Stevens, E.~D.} (eds.)
  \emph{\bibinfo{booktitle}{Tuna: physiology, ecology, and evolution}},
  vol.~\bibinfo{volume}{19}
  (\bibinfo{publisher}{Gulf Professional Publishing},
  \bibinfo{year}{2001}).

\bibitem{kitagawa2015biology}
\bibinfo{editor}{Kitagawa, T.} \& \bibinfo{author}{Kimura, S.} (eds.)
  \emph{\bibinfo{booktitle}{Biology and ecology of bluefin tuna}}
  (\bibinfo{publisher}{CRC Press},
  \bibinfo{year}{2015}).

\bibitem{llopiz2015global}
\bibinfo{author}{Llopiz, J.~K.} \& \bibinfo{author}{Hobday, A.~J.}
\newblock \bibinfo{journal}{\bibinfo{title}{A global comparative analysis of
  the feeding dynamics and environmental conditions of larval tunas, mackerels,
  and billfishes}}.
\newblock {\emph{\JournalTitle{Deep Sea Res. Part II Top. Stud. Oceanogr.}}}
  \textbf{\bibinfo{volume}{113}}, \bibinfo{pages}{113--124}
  (\bibinfo{year}{2015}).

\bibitem{reglero2014worldwide}
\bibinfo{author}{Reglero, P.}, \bibinfo{author}{Tittensor, D.~P.},
  \bibinfo{author}{{\'A}lvarez-Berastegui, D.},
  \bibinfo{author}{Aparicio-Gonz{\'a}lez, A.} \& \bibinfo{author}{Worm, B.}
\newblock \bibinfo{journal}{\bibinfo{title}{Worldwide distributions of tuna
  larvae: revisiting hypotheses on environmental requirements for spawning
  habitats}}.
\newblock {\emph{\JournalTitle{Mar. Ecol. Prog. Ser.}}}
  \textbf{\bibinfo{volume}{501}}, \bibinfo{pages}{207--224}
  (\bibinfo{year}{2014}).

\bibitem{shelton2011fluctuations}
  \bibinfo{author}{Shelton, A.~O.} \& \bibinfo{author}{Mangel, M.}
\newblock \bibinfo{journal}{\bibinfo{title}{Fluctuations of fish populations
  and the magnifying effects of fishing}}.
\newblock {\emph{\JournalTitle{Proc. Natl. Acad. Sci. USA}}}
  \textbf{\bibinfo{volume}{108}}, \bibinfo{pages}{7075--7080}
  (\bibinfo{year}{2011}).


\bibitem{hane2022natal}
\bibinfo{author}{Hane, Y.}, \bibinfo{author}{Ushikubo, T.},
  \bibinfo{author}{Yokoyama, Y.}, \bibinfo{author}{Miyairi, Y.} \&
  \bibinfo{author}{Kimura, S.}
\newblock \bibinfo{journal}{\bibinfo{title}{Natal origin of {P}acific bluefin tuna
  \textit{Thunnus orientalis} determined by {SIMS} oxygen isotope analysis of
  otoliths}}.
\newblock {\emph{\JournalTitle{PLOS One}}}
  \textbf{\bibinfo{volume}{17}}, \bibinfo{pages}{e0272850}
  (\bibinfo{year}{2022}).

\bibitem{wells2021nursery}
\bibinfo{author}{Wells, R.~D.}, \bibinfo{author}{Quesnell, V.~A.},
  \bibinfo{author}{Humphreys Jr, R.~L.}, \bibinfo{author}{Dewar, H.} \&
  \bibinfo{author}{Rooker, J.~R.} \emph{et~al.}
\newblock \bibinfo{journal}{\bibinfo{title}{Nursery origin and population
  connectivity of swordfish \textit{Xiphias gladius} in the {N}orth {P}acific {O}cean}}.
\newblock {\emph{\JournalTitle{J. Fish Biol.}}}
  \textbf{\bibinfo{volume}{99}}, \bibinfo{pages}{354--363}
  (\bibinfo{year}{2021}).

\bibitem{schaefer2019spatiotemporal}
  \bibinfo{author}{Schaefer, K.~M.} \& \bibinfo{author}{Fuller, D.~W.}
\newblock \bibinfo{journal}{\bibinfo{title}{Spatiotemporal variability in the
  reproductive dynamics of skipjack tuna (\textit{Katsuwonus pelamis}) in the
  eastern Pacific Ocean}}.
\newblock {\emph{\JournalTitle{Fish. Res.}}}
  \textbf{\bibinfo{volume}{209}}, \bibinfo{pages}{1--13}
  (\bibinfo{year}{2019}).

\bibitem{kim2015effect}
  \bibinfo{author}{Kim, Y.~S.}, \bibinfo{author}{Delgado, D.~I.},
  \bibinfo{author}{Cano, I.~A.} \& \bibinfo{author}{Sawada, Y.}
\newblock \bibinfo{journal}{\bibinfo{title}{Effect of temperature and salinity on
  hatching and larval survival of yellowfin tuna (\textit{Thunnus albacares})}}.
\newblock {\emph{\JournalTitle{Fish. Sci.}}}
  \textbf{\bibinfo{volume}{81}}, \bibinfo{pages}{891--897}
  (\bibinfo{year}{2015}).

\bibitem{wexler2011temperature}
  \bibinfo{author}{Wexler, J.~B.}, \bibinfo{author}{Margulies, D.} \&
  \bibinfo{author}{Scholey, V.~P.}
\newblock \bibinfo{journal}{\bibinfo{title}{Temperature and dissolved oxygen
  requirements for survival of yellowfin tuna (\textit{Thunnus albacares}) larvae}}.
\newblock {\emph{\JournalTitle{J. Exp. Mar. Biol. Ecol.}}}
  \textbf{\bibinfo{volume}{404}}, \bibinfo{pages}{63--72}
  (\bibinfo{year}{2011}).

\bibitem{hyde2006central}
\bibinfo{author}{Hyde, J.~R.}, \bibinfo{author}{Humphreys, R.},
  \bibinfo{author}{Musyl, M.}, \bibinfo{author}{Lynn, E.} \&
  \bibinfo{author}{Vetter, R.}
\newblock \bibinfo{journal}{\bibinfo{title}{A central {N}orth {P}acific spawning
  ground for striped marlin \textit{Tetrapturus audax}}}.
\newblock {\emph{\JournalTitle{Bull. Mar. Sci.}}}
  \textbf{\bibinfo{volume}{79}}, \bibinfo{pages}{683--690}
  (\bibinfo{year}{2006}).

\bibitem{fujioka2021habitat}
  \bibinfo{author}{Fujioka, K.}, \bibinfo{author}{Sasagawa, K.},
  \bibinfo{author}{Kuwahara, T.}, \bibinfo{author}{Estess, E.~E.},
  \bibinfo{author}{Takahara, Y.} \emph{et~al.}
\newblock \bibinfo{journal}{\bibinfo{title}{Habitat use of adult Pacific bluefin tuna
  \textit{Thunnus orientalis} during the spawning season in the Sea of Japan: evidence
  for a trade-off between thermal preference and reproductive activity}}.
\newblock {\emph{\JournalTitle{Mar. Ecol. Prog. Ser.}}}
  \textbf{\bibinfo{volume}{668}}, \bibinfo{pages}{1--20}
  (\bibinfo{year}{2021}).


\bibitem{buenafe2022global}
\bibinfo{author}{Buenafe, K. C.~V.}, \bibinfo{author}{Everett, J.~D.},
  \bibinfo{author}{Dunn, D.~C.}, \bibinfo{author}{Mercer, J.},
  \bibinfo{author}{Suthers, I.~M.} \emph{et~al.}
\newblock \bibinfo{journal}{\bibinfo{title}{A global, historical database of
  tuna, billfish, and saury larval distributions}}.
\newblock {\emph{\JournalTitle{Sci. Data}}} \textbf{\bibinfo{volume}{9}},
  \bibinfo{pages}{423} (\bibinfo{year}{2022}).

\bibitem{nishikawa1985average}
\bibinfo{author}{Nishikawa, Y.}
\newblock \bibinfo{title}{Average distribution of larvae of oceanic species of
  scombroid fishes, 1956-1981}.
\newblock \bibinfo{howpublished}{Far Seas Fisheries Research Laboratory}
  (\bibinfo{year}{1985}).
\newblock \bibinfo{note}{Series 12, 99~pp. Archived at: \url{}.}

\bibitem{garcia2020global}
\bibinfo{author}{Garc{\'\i}a~Molinos, J.}
\newblock \bibinfo{journal}{\bibinfo{title}{Global marine warming in a new
  dimension}}.
\newblock {\emph{\JournalTitle{Nat. Ecol. Evol.}}}
  \textbf{\bibinfo{volume}{4}}, \bibinfo{pages}{16--17} (\bibinfo{year}{2020}).

\bibitem{oremus2020governance}
\bibinfo{author}{Oremus, K.~L.} \emph{et~al.}
\newblock \bibinfo{journal}{\bibinfo{title}{Governance challenges for tropical
  nations losing fish species due to climate change}}.
\newblock {\emph{\JournalTitle{Nat. Sustain.}}} \textbf{\bibinfo{volume}{3}},
  \bibinfo{pages}{277--280} (\bibinfo{year}{2020}).

\bibitem{burrows2019ocean}
\bibinfo{author}{Burrows, M.~T.}, \bibinfo{author}{Bates, A.~E.},
  \bibinfo{author}{Costello, M.~J.}, \bibinfo{author}{Edwards, M.},
  \bibinfo{author}{Edgar, G.~J.} \emph{et~al.}
\newblock \bibinfo{journal}{\bibinfo{title}{Ocean community warming responses
  explained by thermal affinities and temperature gradients}}.
\newblock {\emph{\JournalTitle{Nat. Clim. Change}}}
  \textbf{\bibinfo{volume}{9}}, \bibinfo{pages}{959--963}
  (\bibinfo{year}{2019}).

\bibitem{garcia2016climate}
\bibinfo{author}{Garc{\'\i}a~Molinos, J.}, \bibinfo{author}{Halpern, B.~S.},
  \bibinfo{author}{Schoeman, D.~S.}, \bibinfo{author}{Brown, C.J.},
  \bibinfo{author}{Kiessling, W.} \emph{etal.}
\newblock \bibinfo{journal}{\bibinfo{title}{Climate velocity and the future
  global redistribution of marine biodiversity}}.
\newblock {\emph{\JournalTitle{Nat. Clim. Change}}}
  \textbf{\bibinfo{volume}{6}}, \bibinfo{pages}{83--88} (\bibinfo{year}{2016}).

\bibitem{burrows2014geographical}
\bibinfo{author}{Burrows, M.~T.}, \bibinfo{author}{Schoeman, D.~S.}, 
  \bibinfo{author}{Richardson, A.~J.}, \bibinfo{author}{Garc{\'\i}a Molinos, J.},
  \bibinfo{author}{Hoffmann, A.} \emph{et~al.}
\newblock \bibinfo{journal}{\bibinfo{title}{Geographical limits to
  species-range shifts are suggested by climate velocity}}.
\newblock {\emph{\JournalTitle{Nature}}} \textbf{\bibinfo{volume}{507}},
  \bibinfo{pages}{492--495} (\bibinfo{year}{2014}).

\bibitem{reshef2011detecting}
\bibinfo{author}{Reshef, D.~N.}, \bibinfo{author}{Reshef, Y.~A.}, 
  \bibinfo{author}{Finucane, H.~K.}, \bibinfo{author}{Grossman, S.~R.},
  \bibinfo{author}{McVean, G.} \emph{et~al.}
\newblock \bibinfo{journal}{\bibinfo{title}{Detecting novel associations in
  large data sets}}.
\newblock {\emph{\JournalTitle{Science}}} \textbf{\bibinfo{volume}{334}},
  \bibinfo{pages}{1518--1524} (\bibinfo{year}{2011}).

\bibitem{thorson2015spatial}
\bibinfo{author}{Thorson, J.~T.},  \bibinfo{author}{Scheuerell, M.~D.},
  \bibinfo{author}{Shelton, A.~O.}, \bibinfo{author}{See, K.~E.},
  \bibinfo{author}{Skaug, H.~J.} \& \bibinfo{author}{Kristensen, K.}
\newblock \bibinfo{journal}{\bibinfo{title}{Spatial factor analysis: a new tool
  for estimating joint species distributions and correlations in species
  range}}.
\newblock {\emph{\JournalTitle{Methods Ecol. Evol.}}}
  \textbf{\bibinfo{volume}{6}}, \bibinfo{pages}{627--637}
  (\bibinfo{year}{2015}).

\bibitem{liu2019eastern}
\bibinfo{author}{Liu, J.}, \bibinfo{author}{Tian, J.}, \bibinfo{author}{Liu, Z.},
  \bibinfo{author}{Herbert, T.~D.}, \bibinfo{author}{Fedorov, A.V.} \&
  \bibinfo{author}{Lyle, M.} 
\newblock \bibinfo{journal}{\bibinfo{title}{Eastern equatorial {P}acific cold
  tongue evolution since the late {M}iocene linked to extratropical climate}}.
\newblock {\emph{\JournalTitle{Sci. Adv.}}} \textbf{\bibinfo{volume}{5}},
  \bibinfo{pages}{eaau6060} (\bibinfo{year}{2019}).

\bibitem{kang2010eddy}
\bibinfo{author}{Kang, L.}, \bibinfo{author}{Wang, F.} \&
  \bibinfo{author}{Chen, Y.}
\newblock \bibinfo{journal}{\bibinfo{title}{Eddy generation and evolution in
  the {N}orth {P}acific {S}ubtropical {C}ountercurrent ({NPSC}) zone}}.
\newblock {\emph{\JournalTitle{Chin. J. Oceanol. Limnol.}}}
  \textbf{\bibinfo{volume}{28}}, \bibinfo{pages}{968--973}
  (\bibinfo{year}{2010}).

\bibitem{kitagawa2010restricted}
\bibinfo{author}{Kitagawa, T.}, \bibinfo{author}{Kato, Y.},
  \bibinfo{author}{Miller, M.~J.}, \bibinfo{author}{Sasai, Y.},
  \bibinfo{author}{Sasaki, H.} \& \bibinfo{author}{Kimura, S.}
\newblock \bibinfo{journal}{\bibinfo{title}{The restricted spawning area
  and season of {P}acific bluefin tuna facilitate use of nursery areas:
  a modeling approach to larval and juvenile dispersal processes}}.
\newblock {\emph{\JournalTitle{J. Exp. Mar. Biol. Ecol.}}}
  \textbf{\bibinfo{volume}{393}}, \bibinfo{pages}{23--31}
  (\bibinfo{year}{2010}).

\bibitem{fujioka2024influence}
\bibinfo{author}{Fujioka, K.}, \bibinfo{author}{Aoki, Y.},
  \bibinfo{author}{Tsuda, Y.}, \bibinfo{author}{Okamoto, K.},
  \bibinfo{author}{Tsuchida, H.} \emph{et~al.}
\newblock \bibinfo{journal}{\bibinfo{title}{Influence of temperature on hatching
  success of skipjack tuna (\textit{Katsuwonus pelamis}): Implications for spawning
  availability of warm habitats}}.
\newblock {\emph{\JournalTitle{J. Fish Biol.}}}
  \textbf{\bibinfo{volume}{105}}, \bibinfo{pages}{372--377}
  (\bibinfo{year}{2024}).

\bibitem{margulies2007spawning}
\bibinfo{author}{Margulies, D.}, \bibinfo{author}{Sutter, J.~M.},
  \bibinfo{author}{Hunt, S.~L.}, \bibinfo{author}{Olson, R.~J.} \emph{et~al.}
\newblock \bibinfo{journal}{\bibinfo{title}{Spawning and early development
  of captive yellowfin tuna (\textit{Thunnus albacares})}}.
\newblock {\emph{\JournalTitle{Fish. Bull.}}}
  \textbf{\bibinfo{volume}{105}}, \bibinfo{pages}{249--265}
  (\bibinfo{year}{2007}).

\bibitem{pons2017effects}
\bibinfo{author}{Pons, M.}, \bibinfo{author}{Branch, T.~A.},
  \bibinfo{author}{Melnychuk, M.~C.}, \bibinfo{author}{Jensen, O.~P.},
  \bibinfo{author}{Brodziak, J.} \emph{et~al.}
\newblock \bibinfo{journal}{\bibinfo{title}{Effects of biological, economic and
  management factors on tuna and billfish stock status}}.
\newblock {\emph{\JournalTitle{Fish Fish.}}}
  \textbf{\bibinfo{volume}{18}}, \bibinfo{pages}{1--21}
  (\bibinfo{year}{2017}).

\bibitem{hernandez2019evidence}
\bibinfo{author}{Hern\'{a}ndez, C.~M.}, \bibinfo{author}{Witting, J.},
  \bibinfo{author}{Willis, C.}, \bibinfo{author}{Thorrold, S.~R.},
  \bibinfo{author}{Llopiz, J.~K.} \& \bibinfo{author}{Rotjan, R.~D.}
\newblock \bibinfo{journal}{\bibinfo{title}{Evidence and patterns of tuna spawning
  inside a large no-take {M}arine {P}rotected {A}rea}}.
\newblock {\emph{\JournalTitle{Sci. Rep.}}}
  \textbf{\bibinfo{volume}{9}}, \bibinfo{pages}{10772}
  (\bibinfo{year}{2019}).

\bibitem{armas1999confirmation}
  \bibinfo{author}{Gonz\'{a}lez-Armas, R.}, \bibinfo{author}{Sosa-Nishizaki, O.},
  \bibinfo{author}{Funes-Rodr\'{i}guez, R.} \& \bibinfo{author}{Levy-P\'{e}rez, V.~A.}
\newblock \bibinfo{journal}{\bibinfo{title}{Confirmation of the spawning area
  of the striped marlin, \textit{Tetrapturus audax}, in the so-called core area
  of the eastern tropical {P}acific off {M}exico}}.
\newblock {\emph{\JournalTitle{Fish. Oceanogr.}}}
  \textbf{\bibinfo{volume}{8}}, \bibinfo{pages}{238--242}
  (\bibinfo{year}{1999}).

\bibitem{wexler2023effect}
  \bibinfo{author}{Wexler, J.~B.}, \bibinfo{author}{Margulies, D.},
  \bibinfo{author}{Scholey, V.}, \bibinfo{author}{Lennert-Cody, C.~E.},
  \bibinfo{author}{Bromhead, D.} \emph{et~al.}
\newblock \bibinfo{journal}{\bibinfo{title}{The effect of ocean acidification
  on otolith morphology in larvae of a tropical, epipelagic fish species,
  yellowfin tuna (\textit{Thunnus albacares})}}.
\newblock {\emph{\JournalTitle{J. Exp. Mar. Biol. Ecol.}}}
  \textbf{\bibinfo{volume}{569}}, \bibinfo{pages}{151949}
  (\bibinfo{year}{2023}).

\bibitem{frommel2016ocean}
\bibinfo{author}{Frommel, A.~Y.}, \bibinfo{author}{Margulies, D.},
  \bibinfo{author}{Wexler, J.~B.}, \bibinfo{author}{Stein, M.~S.},
  \bibinfo{author}{Scholey, V.~P.} \emph{et~al.}
\newblock \bibinfo{journal}{\bibinfo{title}{Ocean acidification has lethal and sub-lethal
  effects on larval development of yellowfin tuna (\textit{Thunnus albacares})}}.
\newblock {\emph{\JournalTitle{J. Exp. Mar. Biol. Ecol.}}}
  \textbf{\bibinfo{volume}{482}}, \bibinfo{pages}{18--24}
  (\bibinfo{year}{2016}).

\bibitem{jacobson2005studying}
\bibinfo{author}{Jacobson, M.~Z.}
\newblock \bibinfo{journal}{\bibinfo{title}{Studying ocean acidification with
  conservative, stable numerical schemes for nonequilibrium air-ocean exchange
  and ocean equilibrium chemistry}}.
\newblock {\emph{\JournalTitle{J. Geophys. Res. Atmos.}}}
  \textbf{\bibinfo{volume}{110}} (\bibinfo{year}{2005}).

\bibitem{martiny2019biogeochemical}
\bibinfo{author}{Martiny, A.~C.}, \bibinfo{author}{Lomas, M.~W.},
  \bibinfo{author}{Fu, W.}, \bibinfo{author}{Boyd, P.~W.},
  \bibinfo{author}{Chen, Y. L.~L.} \emph{et~al.}
\newblock \bibinfo{journal}{\bibinfo{title}{Biogeochemical controls of surface
  ocean phosphate}}.
\newblock {\emph{\JournalTitle{Sci. Adv.}}} \textbf{\bibinfo{volume}{5}},
  \bibinfo{pages}{eaax0341} (\bibinfo{year}{2019}).

\bibitem{tyrrell1999relative}
\bibinfo{author}{Tyrrell, T.}
\newblock \bibinfo{journal}{\bibinfo{title}{The relative influences of nitrogen
  and phosphorus on oceanic primary production}}.
\newblock {\emph{\JournalTitle{Nature}}} \textbf{\bibinfo{volume}{400}},
  \bibinfo{pages}{525--531} (\bibinfo{year}{1999}).

\bibitem{reglero2011cannibalism}
\bibinfo{author}{Reglero, P.}, \bibinfo{author}{Urtizberea, A.},
  \bibinfo{author}{Torres, A.~P.}, \bibinfo{author}{Alemany, F.} \&
  \bibinfo{author}{Fiksen, {\O}.}
\newblock \bibinfo{journal}{\bibinfo{title}{Cannibalism among size classes of
  larvae may be a substantial mortality component in tuna}}.
\newblock {\emph{\JournalTitle{Mar. Ecol. Prog. Ser.}}}
  \textbf{\bibinfo{volume}{433}}, \bibinfo{pages}{205--219}
  (\bibinfo{year}{2011}).

\bibitem{bakun2003environmental}
\bibinfo{author}{Bakun, A.} \& \bibinfo{author}{Broad, K.}
\newblock \bibinfo{journal}{\bibinfo{title}{Environmental `loopholes' and fish
  population dynamics: comparative pattern recognition with focus on {E}l
  {N}i{\~n}o effects in the {P}acific}}.
\newblock {\emph{\JournalTitle{Fish. Oceanogr.}}}
  \textbf{\bibinfo{volume}{12}}, \bibinfo{pages}{458--473}
  (\bibinfo{year}{2003}).

\bibitem{rutherford1999environmental}
\bibinfo{author}{Rutherford, S.}, \bibinfo{author}{D'Hondt, S.} \&
  \bibinfo{author}{Prell, W.}
\newblock \bibinfo{journal}{\bibinfo{title}{Environmental controls on the geographic
   distribution of zooplankton diversity}}. 
\newblock {\emph{\JournalTitle{Nature}}}
  \textbf{\bibinfo{volume}{400}}, \bibinfo{pages}{749--753}
  (\bibinfo{year}{1999}).

\bibitem{tawa2020fine}
\bibinfo{author}{Tawa, A.}, \bibinfo{author}{Kodama, T.},
  \bibinfo{author}{Sakuma, K.}, \bibinfo{author}{Ishihara, T.} \&
  \bibinfo{author}{Ohshimo, S.}
\newblock \bibinfo{journal}{\bibinfo{title}{Fine-scale horizontal distributions
  of multiple species of larval tuna off the {N}ansei {I}slands, {J}apan}}.
\newblock {\emph{\JournalTitle{Mar. Ecol. Prog. Ser.}}}
  \textbf{\bibinfo{volume}{636}}, \bibinfo{pages}{123--137}
  (\bibinfo{year}{2020}).

\bibitem{rooker2012distribution}
\bibinfo{author}{Rooker, J.~R.}, \bibinfo{author}{Simms, J.~R.},
  \bibinfo{author}{Wells, R.~D.}, \bibinfo{author}{Holt, S.~A.},
  \bibinfo{author}{Holt, G.~J.} \emph{et~al.}
\newblock \bibinfo{journal}{\bibinfo{title}{Distribution and habitat
  associations of billfish and swordfish larvae across mesoscale features in
  the {G}ulf of {M}exico}}.
\newblock {\emph{\JournalTitle{PLOS ONE}}} \textbf{\bibinfo{volume}{7}},
  \bibinfo{pages}{e34180} (\bibinfo{year}{2012}).

\bibitem{warton2015so}
\bibinfo{author}{Warton, D.~I.}, \bibinfo{author}{Blanchet, F.~G.},
  \bibinfo{author}{O'Hara, R.~B.}, \bibinfo{author}{Ovaskainen, O.},
  \bibinfo{author}{Taskinen, S.} \emph{et~al.}
\newblock \bibinfo{journal}{\bibinfo{title}{So many variables: joint
  modeling in community ecology}}.
\newblock {\emph{\JournalTitle{Trends Ecol. Evol.}}} \textbf{\bibinfo{volume}{30}},
  \bibinfo{pages}{766-779} (\bibinfo{year}{2015}).

\bibitem{cameletti2013spatio}
\bibinfo{author}{Cameletti, M.}, \bibinfo{author}{Lindgren, F.},
  \bibinfo{author}{Simpson, D.} \& \bibinfo{author}{Rue, H.}
\newblock \bibinfo{journal}{\bibinfo{title}{Spatio-temporal modeling of
  particulate matter concentration through the {SPDE} approach}}.
\newblock {\emph{\JournalTitle{AStA Adv. Stat. Anal.}}}
  \textbf{\bibinfo{volume}{97}}, \bibinfo{pages}{109--131}
  (\bibinfo{year}{2013}).

\bibitem{kristensen2016tmb}
\bibinfo{author}{Kristensen, K.}, \bibinfo{author}{Nielsen, A.},
  \bibinfo{author}{Berg, C.~W.}, \bibinfo{author}{Skaug, H.} \&
  \bibinfo{author}{Bell, B.~M.}
\newblock \bibinfo{journal}{\bibinfo{title}{{TMB}: Automatic differentiation
  and {L}aplace approximation}}.
\newblock {\emph{\JournalTitle{J. Stat. Softw.}}}
  \textbf{\bibinfo{volume}{70}}, \bibinfo{pages}{1--21} (\bibinfo{year}{2016}).

\bibitem{wood2020inference}
\bibinfo{author}{Wood, S.~N.}
\newblock \bibinfo{journal}{\bibinfo{title}{Inference and computation with
  generalized additive models and their extensions}}.
\newblock {\emph{\JournalTitle{Test}}}
  \textbf{\bibinfo{volume}{29}}, \bibinfo{pages}{307--339} (\bibinfo{year}{2020}).

\bibitem{wang1998scale}
\bibinfo{author}{Wang, Y.-P.} \& \bibinfo{author}{Lee, S.~L.}
\newblock \bibinfo{journal}{\bibinfo{title}{Scale-space derived from {B}-splines}}.
\newblock {\emph{\JournalTitle{IEEE Trans. Pattern Anal. Mach. Intell.}}}
  \textbf{\bibinfo{volume}{20}}, \bibinfo{pages}{1040--1055} (\bibinfo{year}{1998}).


\end{thebibliography}

\paragraph*{Acknowledgements.} We are grateful to Hidetada Kiyofuji and Taketoshi Kodama for comments and insights. We acknowledge support from Research and assessment program for internationally managed fisheries resources, the Fisheries Agency of Japan. M.\,J. was partly supported by the Japan Society for the Promotion of Science (JSPS) KAKENHI grant 21H03625.
\paragraph*{Author contributions.} H.\,I. and M.\,J. conceptualised research, interpreted results, and discussed implications. H.\,I. performed modelling work. M.\,J. wrote the manuscript.

\clearpage

\end{document}


\baselineskip24pt

\maketitle 

\renewcommand{\tablename}{Supporting Table}
\renewcommand{\figurename}{Supporting Figure}

\clearpage

\begin{table*}
\caption{Model selection. We hypothesised and tested 14 models of increasing complexity, expressing their predictive power with the mean square and mean absolute errors (MSE and MAE, respectively) obtained via 10-fold cross-validation. Model complexity increased by adding more predictor variables, more latent predictors, or more auto-regressive terms. To maximise predictive power while avoiding overfitting, we chose the model with the lowest MSE and MAE.}
\footnotesize	
\begin{center}
\begin{threeparttable}
\begin{tabular}{c c c c c c c c c}
\hline
No. & Timestep & Predictors\tnote{a} & Features\tnote{b} & Latent\tnote{c} & AR terms\tnote{d} & MSE\tnote{e} & MAE\tnote{f} \\
\hline 
1                   & annual    & $\mathrm{SST}$                  & quad & 1 & 1 &          1.2490 &          0.0875 \\
2                   & annual    & $\mathrm{SST}$                  & quad & 2 & 1 &          1.2260 &          0.0841 \\
3                   & quarterly & $\mathrm{SST}$                  & quad & 2 & 1 &          1.2460 &          0.0825 \\
4                   & quarterly & $\mathrm{SST}$                  & quad & 2 & 4 &          1.2390 &          0.0823 \\
5                   & quarterly & $\mathrm{SST}$, $\mathrm{vTYP}$ & quad & 2 & 4 &          1.2590 &          0.0821 \\
6                   & quarterly & $\mathrm{SST}$, $\mathrm{yr}$   & quad & 2 & 4 &          1.2250 &          0.0804 \\
7                   & quarterly & $\mathrm{SST}$                  & quad & 3 & 4 &          1.2190 &          0.0796 \\
8                   & quarterly & $\mathrm{SST}$, $\mathrm{yr}$   & quad & 3 & 4 &          1.2200 &          0.0790 \\
9                   & quarterly & $\mathrm{SST}$                  & quad & 4 & 4 &          1.2140 &          0.0786 \\
10                  & quarterly & $\mathrm{SST}$, $\mathrm{yr}$   & quad & 4 & 4 &          1.2170 &          0.0786 \\
11                  & quarterly & $\mathrm{SST}$                  & quad & 5 & 4 &          1.2170 &          0.0783 \\
12                  & quarterly & $\mathrm{SST}$                  & sp   & 2 & 4 &          1.2360 &          0.0815 \\
13                  & quarterly & $\mathrm{SST}$                  & sp   & 3 & 4 &          1.2160 &          0.0792 \\
\smash{14\tnote{g}} & quarterly & $\mathrm{SST}$                  & sp   & 4 & 4 & \textbf{1.2130} & \textbf{0.0780} \\
\hline    
\end{tabular}
\begin{tablenotes}
\item[a] Fixed-effect predictor variables: $\mathrm{SST}$ = sea-surface temperature (continuous), $\mathrm{vTYP}$ = vessel type (categorical), and $\mathrm{yr}$ = year (categorical)
\item[a] Feature map: quad = quadratic, sp = spline
\item[c] Number of latent spatial predictors
\item[d] Number of auto-regressive coefficients
\item[r] Mean square error of 10-fold cross-validation
\item[f] Mean absolute error of 10-fold cross-validation
\item[g] Selected model
\end{tablenotes}
\end{threeparttable}
\end{center}
\label{tab:modelselection}
\end{table*}

\clearpage

\begin{figure*}
  \centering
  \includegraphics[width=1.0\textwidth]{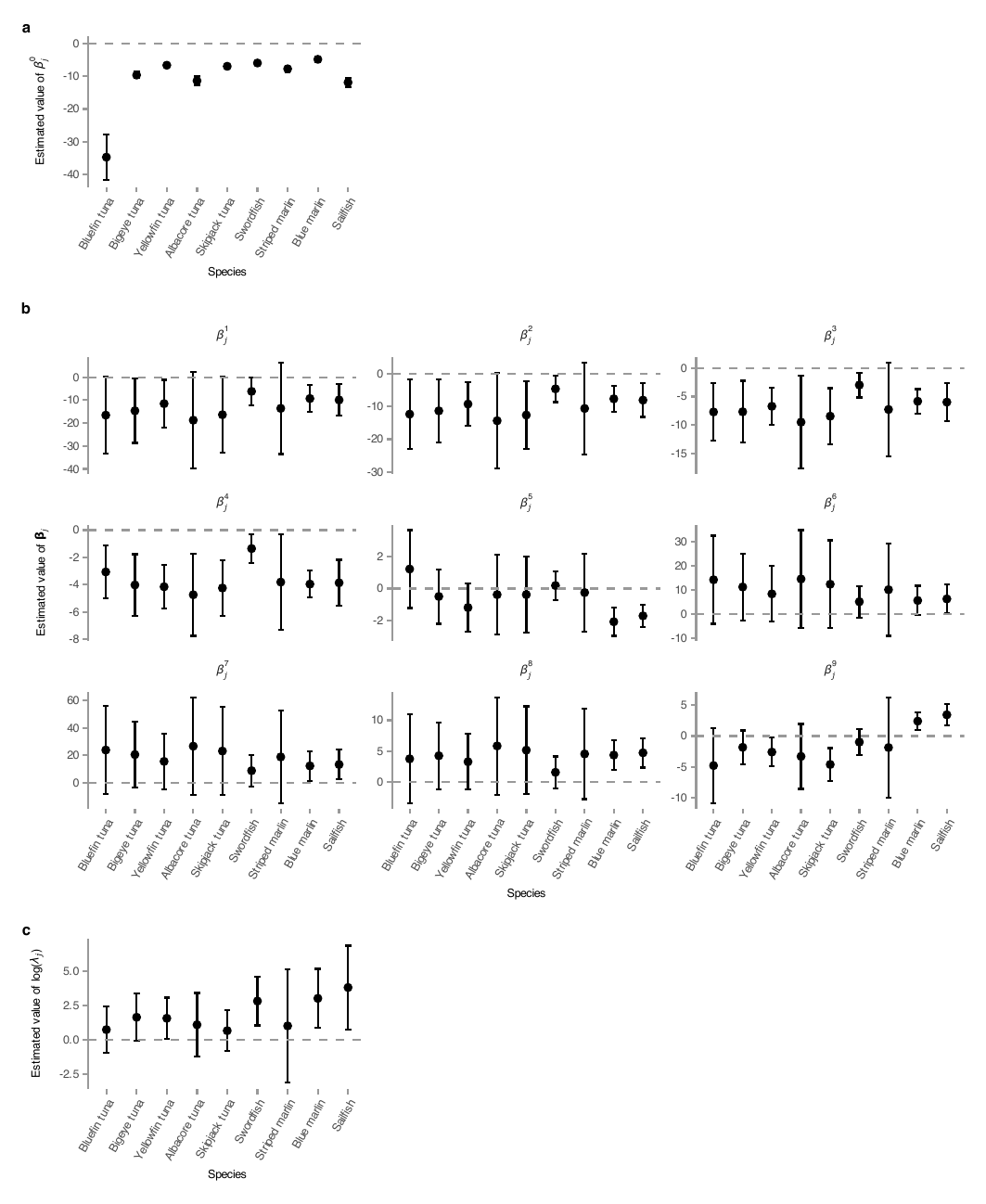}
  \caption{Estimated model intercepts and quantities related to the sea-surface temperature (SST). (b) Intercepts $\beta_j^0$ ($j=1,\ldots,9$). (b) Spline parameters $\beta_j^1,\ldots,\beta_j^9$. (c) Smoothness penalties $\lambda_j$. Circles indicate numerical values and error bars the 95\,\% confidence intervals.}
  \label{fig:fittedparams1}
\end{figure*}

\clearpage

\begin{figure*}
  \centering
  \includegraphics[width=1.0\textwidth]{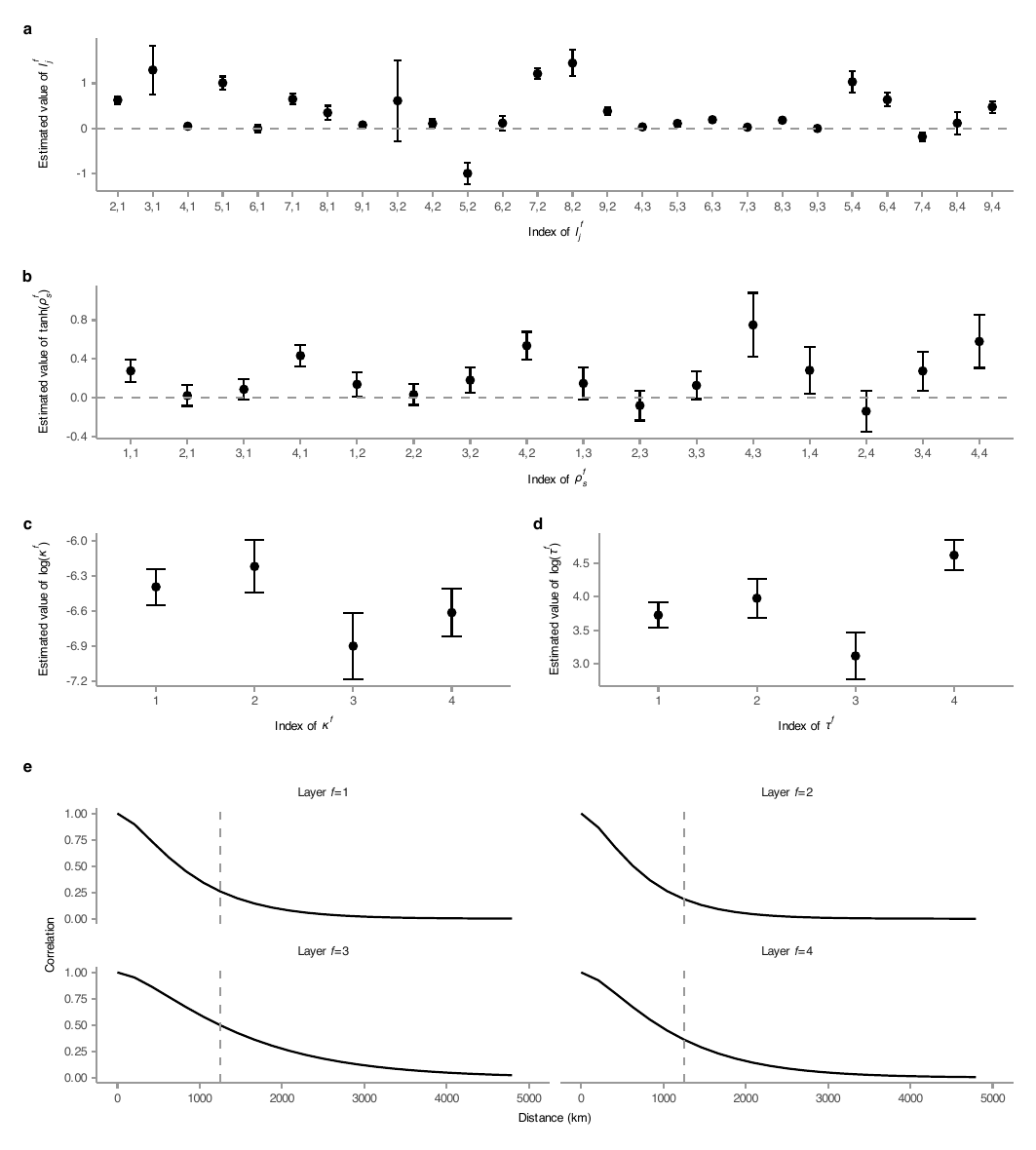}
  \caption{Estimated quantities related to latent predictors. (a) Parameters $l_j^f$ ($j=f+1,\ldots,9$ and $f=1,\ldots,4$). (b) Temporal auto-correlation ($\rho_s^f$, $s=1,\ldots,4$). All significant values are positive, suggesting that a good season in a given year is likely to be followed by another good season in the next year. Likewise for bad seasons. (c,\,d) Spatial auto-correlation ($\kappa^f$ and $\tau^f$). (e) The Mat\'{e}rn covariance function for each of the four latent predictors. The covariance extends beyond the maximum edge distance in the Delaunay triangulation of the study area (vertical dashed lines), thus passing an important model diagnostic. Circles indicate numerical values and error bars the 95\,\% confidence intervals.}
  \label{fig:fittedparams234}
\end{figure*}

\clearpage

\begin{figure*}
  \centering
  \includegraphics[width=1.0\textwidth]{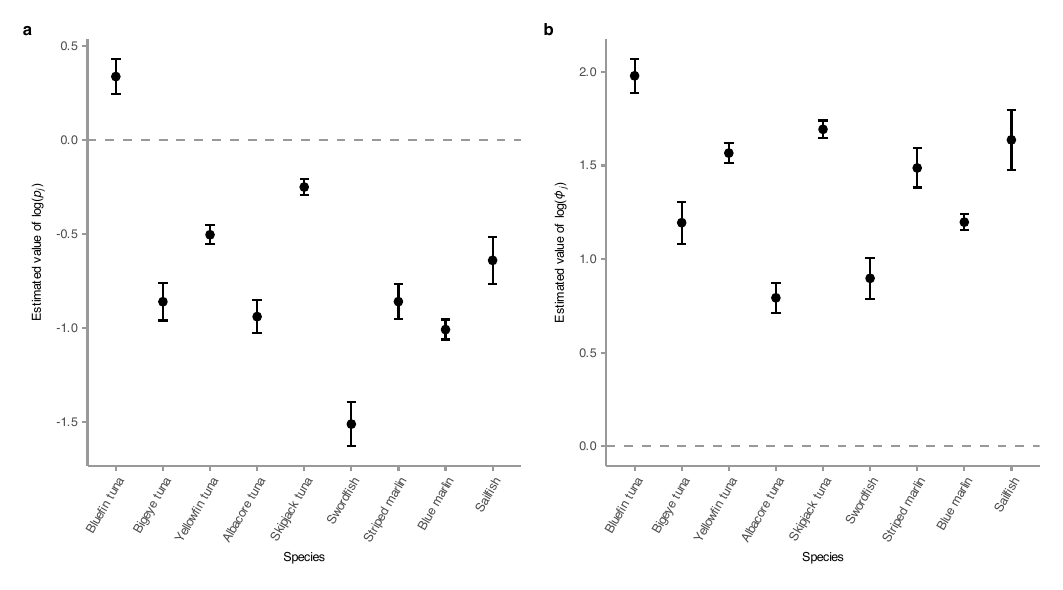}
  \caption{Estimated quantities related to the Tweedie distribution. Circles indicate numerical values and error bars the 95\,\% confidence intervals.}
  \label{fig:fittedparams5}
\end{figure*}

\clearpage

\begin{figure*}
  \centering
  \includegraphics[width=1\textwidth]{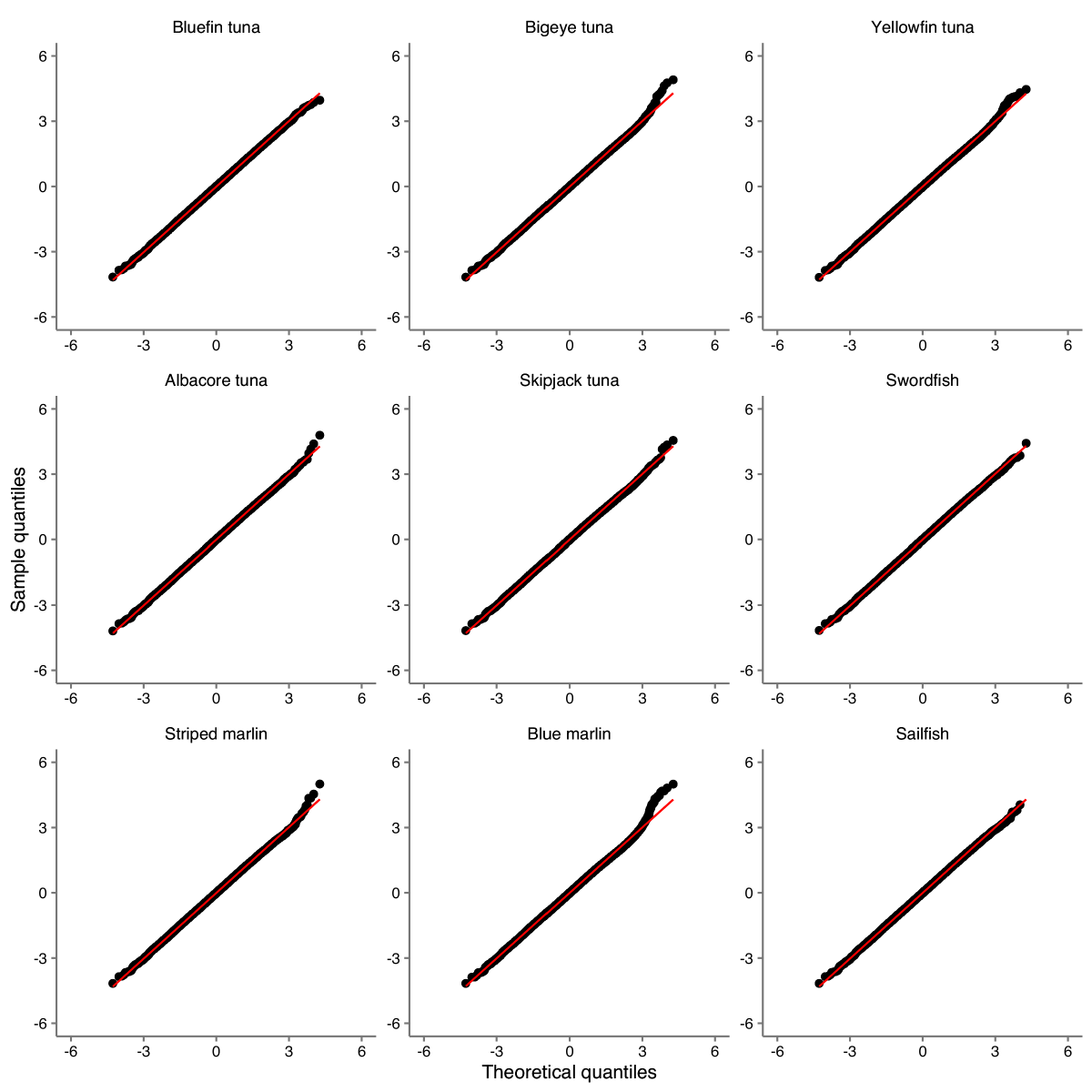}
  \caption{Model diagnostics. We calculated randomised quantile residuals (RQRs) because they offer good statistical power and low type I error in detecting misspecification of count regression models, including non-linear covariate effects, over-dispersion, zero inflation, and even misspecified distributional assumptions~\cite{feng2020comparison}. If our model were misspecified, RQRs in the shown Q-Q plots would deviate substantially from the diagonals shown in red. Lack of such deviations implies that the model satisfactorily passed the diagnostic testing against common misspecifications.}
  \label{fig:diagnostics}
\end{figure*}

\clearpage

\begin{figure*}
  \centering
  \includegraphics[width=1.0\textwidth]{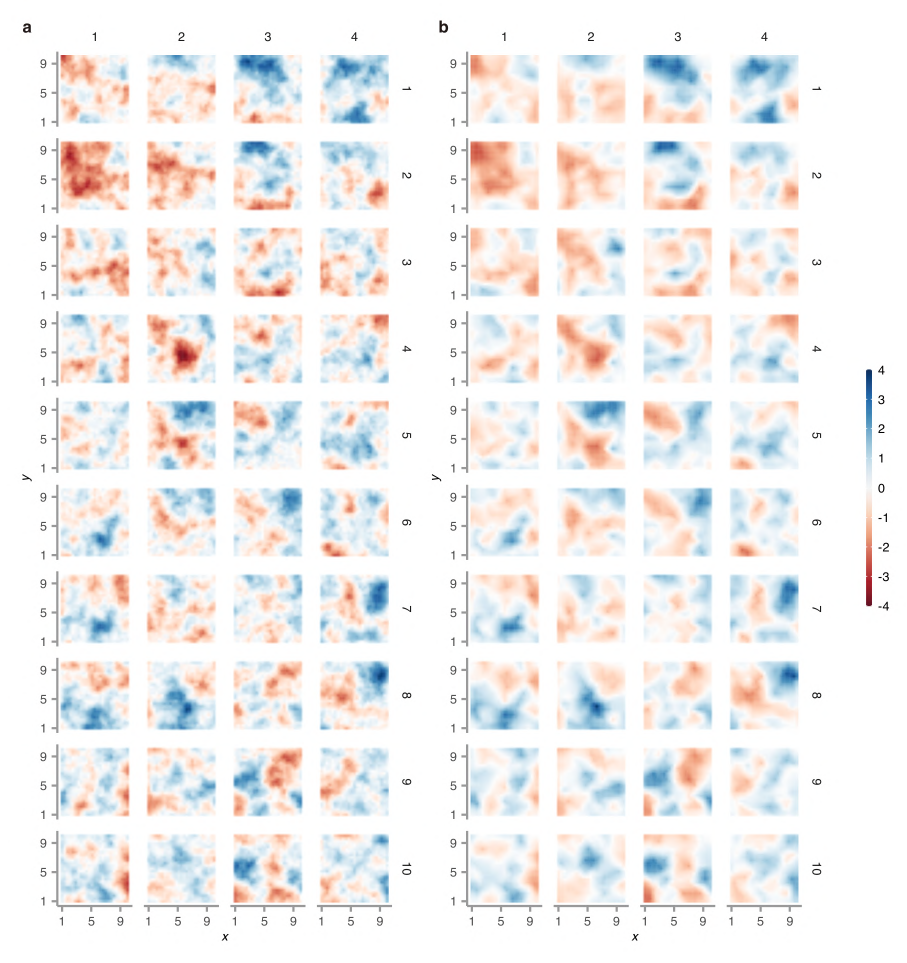}
  \caption{Model diagnostics on an artificial dataset I. We created a 4-season, 10-year long `dataset' by running the model in a configuration with two categorical and two latent predictors. We then randomly dropped 50\,\% of the data points and fitted the model to the remaining data. (a) The `ground-truth' first latent predictor. (b) The model-estimated first latent predictor successfully reproduces the features of its ground-truth counterpart. The $x$ and $y$ coordinates are in an arbitrary unit of distance. See also Supporting Figure~\ref{fig:diagnostics2}.}
  \label{fig:diagnostics1}
\end{figure*}

\clearpage

\begin{figure*}
  \centering
  \includegraphics[width=1.0\textwidth]{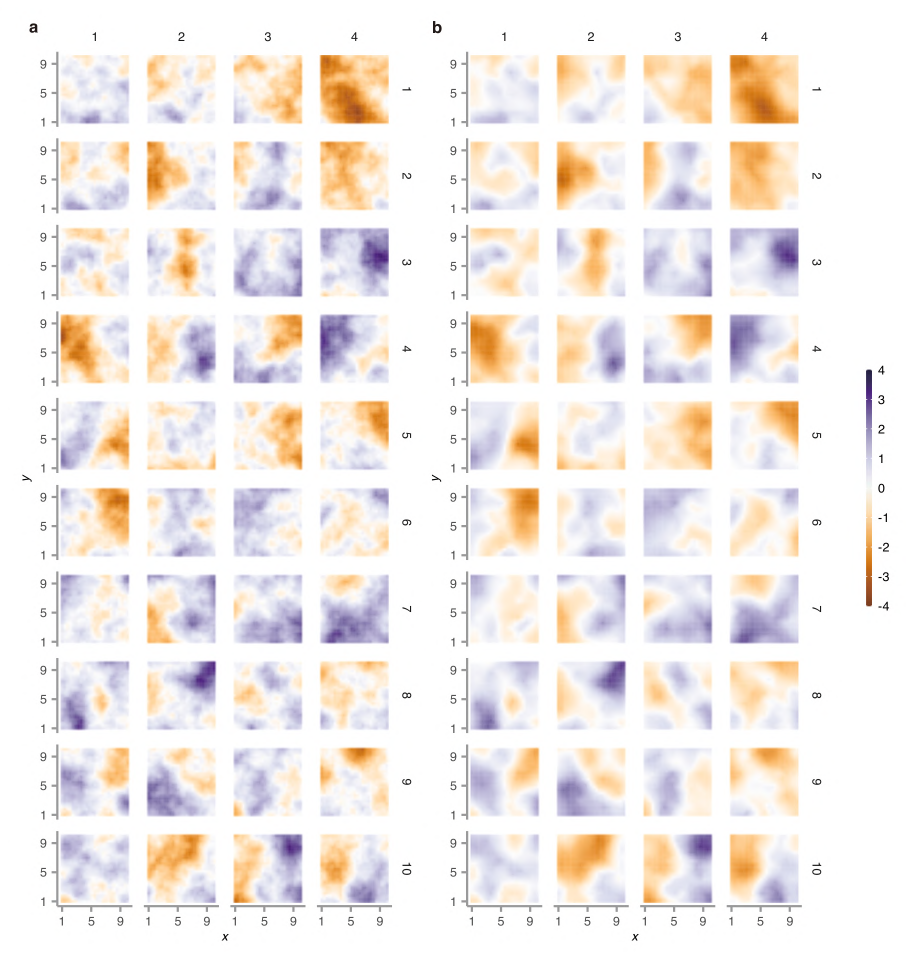}
  \caption{Model diagnostics on an artificial dataset II. The same as Supporting Figure~\ref{fig:diagnostics1}, but for the second latent predictor.}
  \label{fig:diagnostics2}
\end{figure*}

\clearpage

\begin{figure*}
  \centering
  \includegraphics[width=1.0\textwidth]{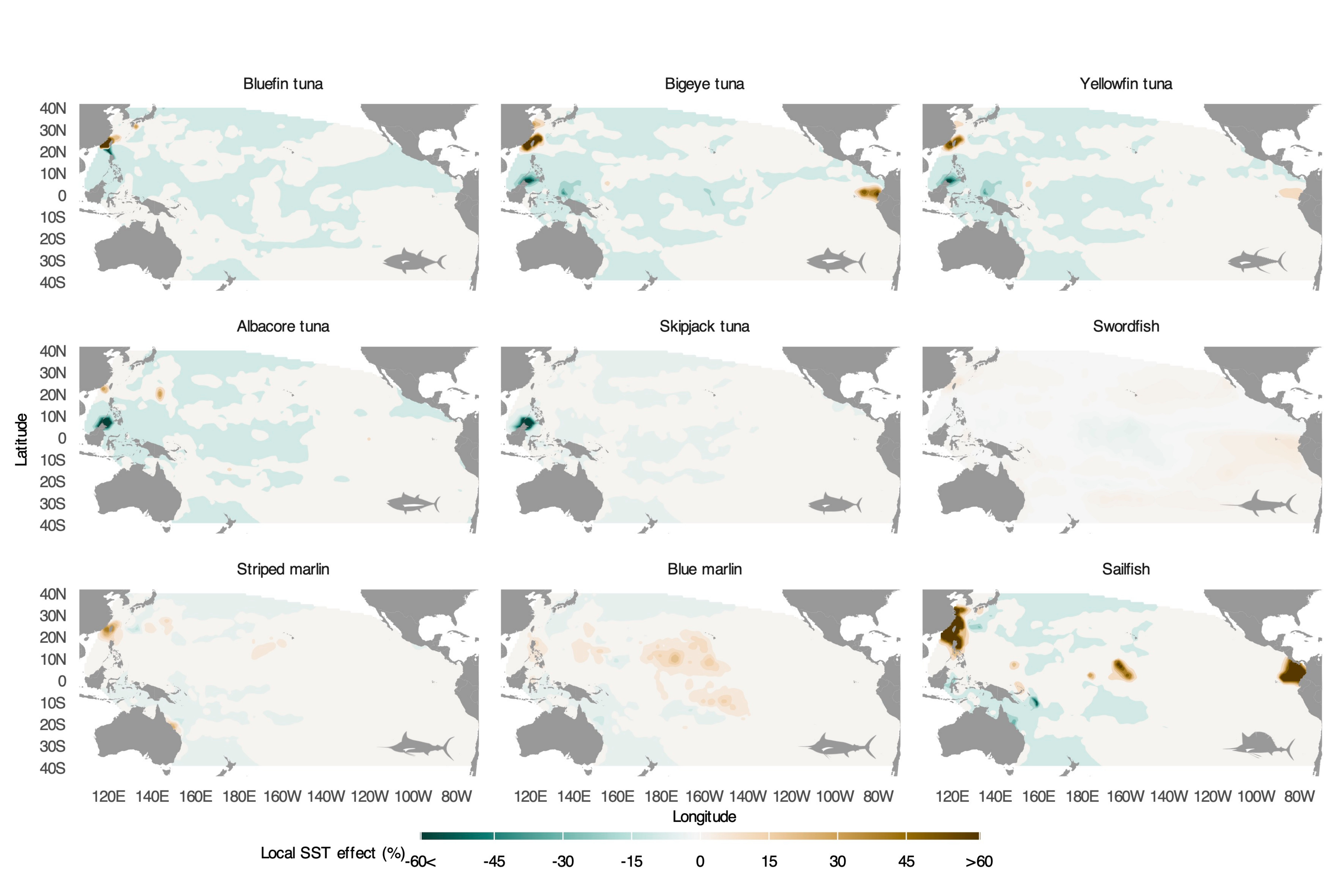}
  \caption{Geospatially resolved predictions in terms of the local expected percentage change in larval density for the El Ni\~{n}o year. Brown and green shades respectively indicate areas where we can expect more or fewer larvae. The same predictions for the current conditions are shown in Figure~\ref{fig:spatiallyresolved}.}
  \label{fig:spatiallyresolved2}
\end{figure*}

\clearpage

\begin{figure*}
  \centering
  \includegraphics[width=1.0\textwidth]{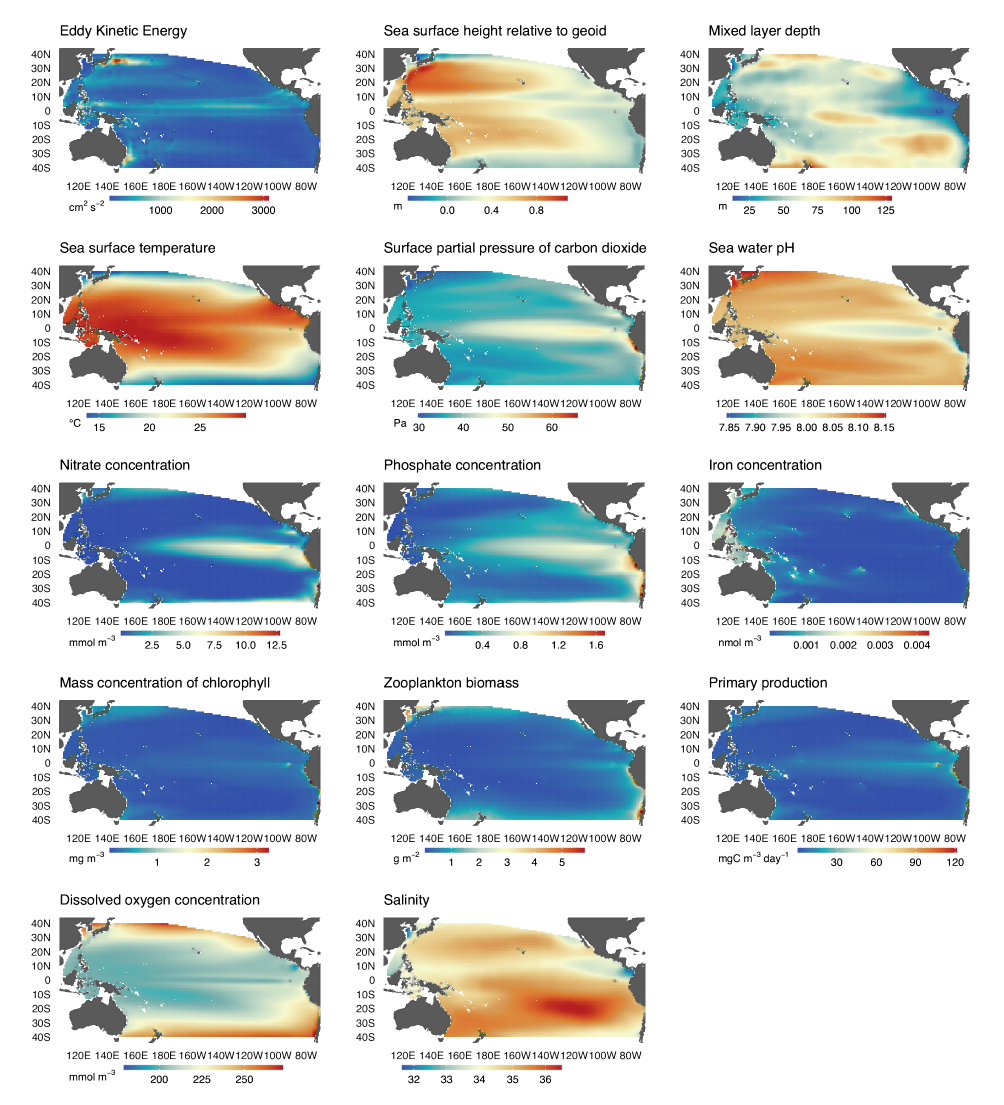}
  \caption{Biogeochemical factors analysed for their association with the oceanic environment's potential to host tuna and billfish larvae. All data is publicly available from sources listed in Methods, Data availablity.}
  \label{fig:biogeochemdata}
\end{figure*}

\clearpage

\begin{figure*}
  \centering
  \includegraphics[width=1.0\textwidth]{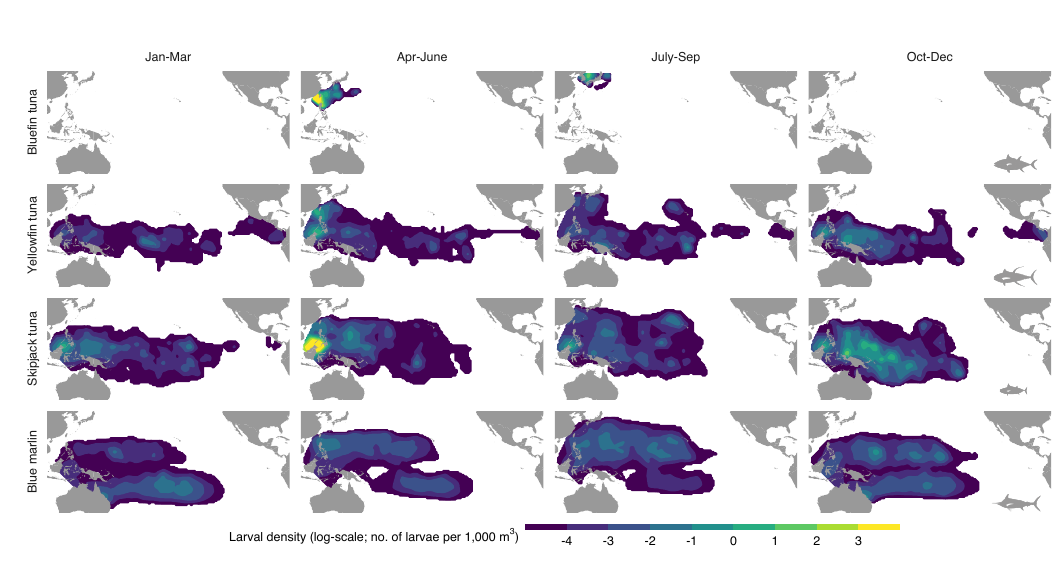}
  \caption{Model sensitivity to input data. Shown are the expected larval distributions for the reference period 1960-1985 when only the four most heavily sampled species are retained as inputs. The results compare favourably to those in Figure~\ref{fig:larvaldensities}. Additional sensitivity runs in which single-species input data is omitted are accessible at \protect\url{https://doi.org/10.17605/OSF.IO/42HM8}.}
  \label{fig:sensitivity}
\end{figure*}

\clearpage

\begin{figure*}
  \centering
  \includegraphics[width=1.0\textwidth]{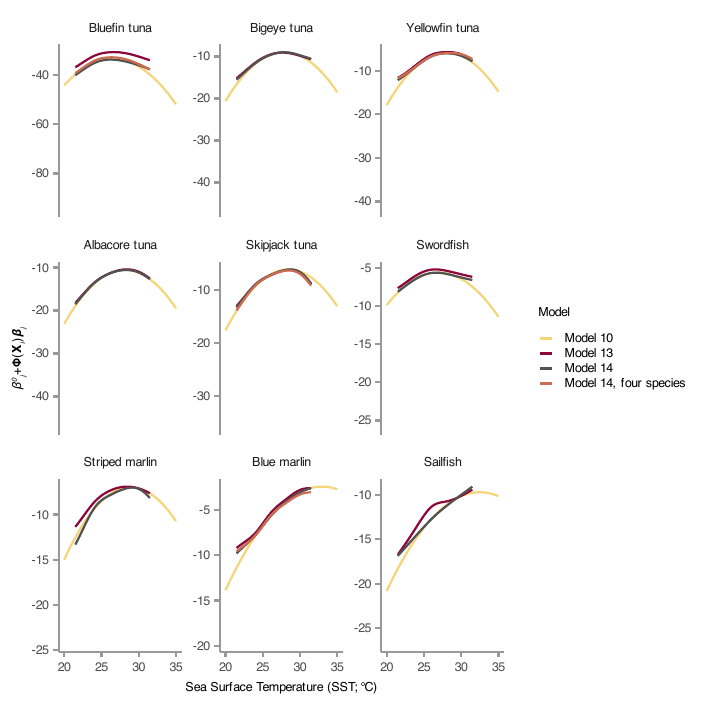}
  \caption{Robustness of temperature-response curves. Overlapping curves are obtained irrespective of whether splines are replaced with parabolas (model 10), three latent predictors are used instead of four (model 13), or input data for some species is omitted (model 14, four species). Unlike splines, parabolas can be extrapolated out-of-sample and are guaranteed to have a maximum that should correspond to the optimal SST for a given species. Such optima for blue marlin and sailfish, however, exceed 30\,$^\circ$C, and defy biological realism. Precise model definitions are given in Supporting Table~\ref{tab:modelselection}.}
  \label{fig:sensitivity2}
\end{figure*}

\clearpage
